\documentclass{article}

\usepackage{diagbox}

\usepackage{booktabs}
\usepackage{pifont} 

\usepackage{ulem}
\usepackage{jheppub}

\usepackage{pdflscape}
\usepackage{float,lscape}
\usepackage{youngtab}

\usepackage{rotating}
\newcommand{\comments}[1]{}

\usepackage{graphicx} 
\usepackage{dcolumn}
\usepackage[T1]{fontenc}
\usepackage[dvipsnames,table]{xcolor}   

\usepackage{arydshln}
\usepackage{mathrsfs}                   

\usepackage{multirow}                   
\usepackage{lipsum}                     
\usepackage{longtable}                  
\usepackage{pifont}                     
\usepackage{slashed}
\usepackage{comment}

\usepackage[compat=1.1.0]{tikz-feynhand}
\usepackage{tikz, pgfplots}
\pgfplotsset{compat=1.18}

\usepackage[detect-all]{siunitx}
\newcommand{\braket}[1]{\ensuremath{\left\langle#1\right\rangle}}

\newcommand{\cmark}{\textcolor{green!60!black}{\ding{51}}}
\newcommand{\xmark}{\textcolor{red!70!black}{\ding{55}}}

\newcommand{\tr}{{\rm Tr}}

\def\simgt{\rlap{\lower 3.5 pt\hbox{$\mathchar \sim$}}\raise 1pt \hbox {$>$}}
\def\simlt{\rlap{\lower 3.5 pt\hbox{$\mathchar \sim$}}\raise 1pt \hbox {$<$}}

\title{
New Avenues for \boldmath $|\Delta B|=2$ Processes Beyond Neutron–Antineutron Oscillations
}

\author[a,b]{Arnau Bas i Beneito,}
\affiliation[a]{Departament de Física Teòrica, Universitat de València, 46100 Burjassot, Spain}
\emailAdd{arnau.bas@ific.uv.es}
\affiliation[b]{Institut de Física Corpuscular (CSIC-Universitat de València),
Parc Científic UV, C/Catedrático José Beltrán, 2, E-46980 Paterna, Spain}

\author[c,d]{Svjetlana Fajfer,} 
\affiliation[c]{Department of Physics, University of Ljubljana, Jadranska 19, 1000 Ljubljana, Slovenia}
\affiliation[d]{Jo\v zef Stefan Institute, Jamova 39, 1000  Ljubljana, Slovenia}
\emailAdd{svjetlana.fajfer@ijs.si}

\author[e]{Alexey A. Petrov}
\affiliation[e]{Department of Physics and Astronomy, University of South Carolina, Columbia, South Carolina 29208, USA}
\emailAdd{apetrov@sc.edu}

\date{\today}
\preprint{USC-TH-2025-03, FTUV-25-1104.4011}
\abstract{
We explore baryon-number-violating ($|\Delta B| = 2$) processes beyond the well-known neutron-antineutron ($n - \bar{n}$) oscillations, focusing on the $\Lambda - \bar \Lambda$ system. The presence of a strange quark in the $\Lambda$ baryon introduces a new set of six-quark operators roughly of the form $(uds)^2$, which are different from the $(udd)^2$ operators responsible for $n - \bar{n}$ oscillations. Using the Standard Model Effective Field Theory (SMEFT), we classify all dimension-9 operators that cause $|\Delta B|=2$ transitions and study their UV completions mediated by exotic scalar fields with trilinear interactions. We demonstrate that in these models, $\Lambda - \bar \Lambda$ oscillations can occur at tree level, with $n - \bar{n}$ mixing potentially appearing at higher loop levels. We employ a chiral effective theory to constrain the effective mass mixing $\delta m_\Lambda$, deriving bounds from current experimental limits on $n - \bar{n}$ oscillations and dinucleon decays such as $p \,p \to K^+ K^+$. These bounds indicate that $\Lambda - \bar{\Lambda}$ oscillations probe a complementary parameter space, sensitive to baryon-number violation at scales up to $10^2-10^3$ TeV. 
We show that the existing indirect bounds make it challenging to provide a competitive bound on $\delta m_\Lambda$ at BESIII. }

\makeatletter
\gdef\@fpheader{}
\makeatother 
\begin{document} 
\maketitle

\flushbottom

\section{Introduction}

The conservation of baryon number ($B$) and lepton number ($L$) is an accidental global symmetry of the Standard Model (SM). Any observation of processes that violate $B - L$ would therefore serve as a clear indication of physics beyond the Standard Model (BSM) and could play a crucial role in explaining the observed baryon asymmetry of the Universe~\cite{Sakharov:1967dj}. Among the possible baryon-number-violating (BNV) phenomena, transitions with $|\Delta B| = 1$, such as proton decay, have been the subject of extensive theoretical and experimental research. However, no evidence of such processes has been found so far, resulting in strict lower limits on the proton lifetime.

Processes with $|\Delta B| = 2$ provide an alternative way for BNV that does not necessarily involve proton decay. These transitions can originate from different dynamical mechanisms and ultraviolet (UV) completions, thus exploring complementary regions of parameter space. The classic example is neutron--antineutron ($n - \bar{n}$) oscillations, first proposed by Kuzmin in 1970~\cite{Kuzmin:1970nx}, which have been studied within various frameworks, including grand unified theories (GUTs)~\cite{Georgi:1974sy,Fritzsch:1974nn,Pati:1973uk} and baryogenesis models. Experiments so far have set strict bounds on the off-diagonal mass term $\delta m_n$, with the most sensitive limits obtained from the Super-Kamiokande experiment, which corresponds to an oscillation time of $\tau_{n-\bar{n}} > 4.7 \times 10^{8}\,\mathrm{s}$~\cite{Super-Kamiokande:2020bov}. Future facilities like Hyper-Kamiokande~\cite{Hyper-Kamiokande:2018ofw}, DUNE~\cite{DUNE:2020lwj}, and HIBEAM/NNBAR~\cite{Addazi:2020nlz} are expected to enhance sensitivity by several orders of magnitude, especially for free neutron oscillations.

Beyond the neutron system, other neutral baryons may exhibit $|\Delta B| = 2$ transitions with qualitatively distinct phenomenology. The $\Lambda$ baryon, containing one strange quark, provides a natural laboratory to explore strangeness-violating baryon oscillations. The $\Lambda - \bar \Lambda$ system is theoretically analogous to $n - \bar{n}$ oscillations but probes six-quark operators of different flavour. Consequently, $\Lambda - \bar\Lambda$ oscillations may arise from operators that are independent of those inducing neutron--antineutron transitions, allowing for scenarios in which the two phenomena have different dynamical origins. The first proposals to search for such oscillations were presented in~\cite{Luk:2007LambdaOscillation,Kang:2009xt}. The BESIII Collaboration has subsequently performed dedicated searches in $J/\psi \to \Lambda \bar \Lambda$ and $J/\psi \to \bar \Lambda p K^-$ decays, obtaining the most stringent direct bound to date, $\delta m_\Lambda < 2.1 \times 10^{-18}\,\mathrm{GeV}$, corresponding to an oscillation time $\tau_{\Lambda-\bar \Lambda} > 1.4 \times 10^{-7}\,\mathrm{s}$~\cite{BESIII:2024gcd}. 

From the theoretical perspective, $\Lambda - \bar\Lambda$ oscillations open new directions for studying BNV processes with quarks of the second generation. In the Standard Model Effective Field Theory (SMEFT) framework, these arise as dimension-9 operators suppressed by powers of a high mass scale $\Lambda_{\mathrm{BNV}}$. UV completions of such operators often involve scalar diquarks or leptoquarks (LQs) with trilinear interactions~\cite {Arnold:2012sd,Goity:1994dq,Babu:2012vc}, which can generate $|\Delta B| = 2$ transitions without inducing $|\Delta B| = 1$ processes such as proton decay. The classification of the relevant operators and their renormalisation-group evolution for $n - \bar n$ oscillations was first presented in Ref.~\cite{Buchoff:2015qwa} and has recently been revisited in Ref.~\cite{ThomasArun:2025rgx}, providing a systematic foundation for analyzing $\Lambda - \bar\Lambda$ phenomenology. Furthermore, the embedding of these $|\Delta B| = 2$ processes within the $\mathrm{SU}(5)$ GUT framework has been thoroughly investigated in Ref.~\cite{deGouvea:2014lva}, while their possible link to the dynamical origin of the matter--antimatter asymmetry of the Universe has been explored in Refs.~\cite{Fridell:2021gag,Babu:2012vc,Herrmann:2014fha,Baldes:2011mh,Babu:2006xc,Babu:2008rq,Babu:2013yca}, and in Ref.~\cite{Bittar:2024nrn}, where a model of baryogenesis was first connected to $\Lambda - \bar \Lambda$ oscillations. The complementarity of different experimental approaches to $n-\bar n$ oscillations has been recently emphasized in Ref.~\cite{Barrow:2025rhm}, motivating a unified treatment of all $|\Delta B| = 2$ phenomena within a common phenomenological framework, in line with the goals of the present work.

In this paper, we explore the phenomenology of BNV processes with $|\Delta B| = 2$ that extend beyond the $n - \bar n$ sector. We develop an effective field theory framework encompassing both neutron and hyperon oscillations and identify the corresponding sets of six-quark operators, along with their scalar-mediated UV completions. We show that available experimental data from $n - \bar{n}$ oscillations and dinucleon decays $p \,p \to K^+ K^+$ put stringent constraints on the parameters of $\Lambda - \bar\Lambda$ oscillations, making it challenging to obtain competitive results at BESIII and LHC.

The paper is organised as follows. In Sec.~\ref{sec:theoreticalframework} we review the key aspects of $|\Delta B|=2$ processes, including hadronic transitions, the EFT framework of the relevant operators, and their purely scalar tree-level UV completions. In Sec.~\ref{sec:exoticpheno} we discuss exotic phenomenology associated to the $|\Delta B|=2$ sector and derive bounds on the parameter $\delta m_{\Lambda}$ from current experimental limits. Sec.~\ref{sec:uvmodel} illustrates these results within a specific UV model. Finally, we summarise our conclusions in Sec.~\ref{sec:conclusions}. The appendices~\ref{sec:fermionscalartopology} and \ref{sec:4scalars} provide additional material on UV completions with heavy fermions and quartic scalar interactions.

\section{General framework for \texorpdfstring{\boldmath $|\Delta B|=2$}{Delta B =2} processes} \label{sec:theoreticalframework}

In this section, we review the fundamental concepts underlying neutral baryon oscillations, following the standard formalism used to study meson systems such as $K^0-\bar K^0$, $D^0- \bar D^0$, and  $B^0_{d,s}-\bar B^0_{d,s}$ oscillations. This framework has been extensively studied for the case of $n-\bar n$ oscillations, but here we extend it in a straightforward way to describe any neutral baryon $\mathcal{B}$. Special focus will be on the lightest electrically neutral baryons $n$ and $\Lambda$.

\subsection{\texorpdfstring{$|\Delta B|=2$}{Delta B =2} hadronic transitions} \label{sec:HeffB2}

The time evolution of $\mathcal{B} - \mathcal{\bar B}$ oscillations is governed by a Schr\"odinger-like equation
\begin{equation}
    i \frac{\partial}{\partial t}
    \begin{pmatrix}
        \mathcal{B} \\ \mathcal{\bar B}
    \end{pmatrix}
    = \mathcal{H}_{\rm eff} \begin{pmatrix}
        \mathcal{B} \\ \mathcal{\bar B}
    \end{pmatrix} \; ,
\end{equation}
where $\mathcal{H}_{\rm eff}$ can be written as \cite{Mohapatra:1980de,Cowsik:1980np,Kuo:1980ew} (see Refs.~\cite{Phillips:2014fgb,Addazi:2020nlz,Mohapatra:2009wp} for reviews)
\begin{equation}
    \label{eq:HeffB2}
    \mathcal{H}_{\rm eff} =
    \begin{pmatrix}
        M_1 & \delta m\\
        \delta m & M_2
    \end{pmatrix} \; .
\end{equation} 
Here $M_i = m_{\cal B} - i\lambda/2 + \alpha_i$, where $m_{\cal B}$ is the mass, $\lambda^{-1} = \tau_{\mathcal{B}}$ is the mean lifetime of a free baryon $\mathcal{B}$, and $\alpha_i$ parametrizes the baryon's interaction with the environment, such as the external magnetic field. While the mass and the width of the neutral baryon and its antiparticle are required to be the same by the CPT symmetry, $\alpha_1$ could be different from $\alpha_2$.

One may diagonalize the Hamiltonian of the two-state quantum system of Eq.~\eqref{eq:HeffB2} through an $SO(2)$ rotation to go to the mass basis, characterized by an angle $\theta$ given by
\begin{equation}
    \tan (2 \theta) = \frac{2\, \delta m}{\Delta M} \; ,
\end{equation}
with $\Delta M = M_1-M_2$, where the real energy eigenvalues are given by
\begin{equation} \label{eq:eigenvalues}
    E_{1,2} = \frac{1}{2}\left[M_1 + M_1 \pm \sqrt{(\Delta M)^2 + 4(\delta m)^2} \right] \,.
\end{equation}
This means that there is a non-zero probability that a neutral Baryon $\mathcal{B}$ converts into its antiparticle $\mathcal{\bar B}$, with a probability given by
\begin{equation} \label{eq:generalBBO}
    P(\mathcal{B} \to \mathcal{\bar B}) = \sin^2(2\theta) \sin^2(\omega \, t) e^{-\lambda t} = \frac{(\delta m)^2}{(\delta m)^2 + (\Delta M/2)^2} \;\sin^2 (\sqrt{(\delta m)^2 + (\Delta M/2)^2} \; t) \,e^{-\lambda t}\; ,
\end{equation}
with $\omega = \sqrt{(\Delta M/2)^2 + (\delta m)^2}$. Free $\mathcal{B} - \mathcal{\bar B}$ oscillations correspond to the case $\Delta M=0$. The application of an external magnetic field splits $M_1$ and $M_2$, usually making $\Delta M \gg \delta m$. In the case of $n$ and $\Lambda$ baryons, experimentally one has $\tau_n = 880$ s, and $\tau_\Lambda = 2.6 \times 10^{-10}$ s~\cite{ParticleDataGroup:2024cfk}. It is important to note that in the limit $\omega t \ll 1$, and $t \ll \tau_{\cal B}$ Eq.~\eqref{eq:generalBBO} can be approximated as
\begin{equation}
P(\mathcal{B} \to \bar{\mathcal{B}}) = (\delta m)^2\; t^2\; e^{-\lambda t} 
\equiv \left( \frac{t}{\tau_{\mathcal{B}-\bar{\mathcal{B}}}} \right)^2 e^{-\lambda t}\, ,
\end{equation}
which is often referred to as the ``quasi-free'' limit. Here the characteristic oscillation time is defined as
\begin{equation}
\tau_{\mathcal{B}-\bar{\mathcal{B}}} = \frac{1}{|\delta m|}\,.
\end{equation}
Hence, a nonzero mass splitting $\delta m$ leads to $\mathcal{B}-\bar{\mathcal{B}}$ oscillations. The formalism applies to free baryons in vacuum, as well as to systems subject to external magnetic fields or, in the case of neutrons, to those bound within nuclei~\cite{Petrov:2021idw,Phillips:2014fgb,Addazi:2020nlz}.

\begin{table}[ht]

\centering
\renewcommand{\arraystretch}{1.5}
\setlength{\tabcolsep}{18pt}

\begin{tabular}{|c|c|c|c|}
\hline
\textbf{ \boldmath $\mathcal{B}$} & \textbf{Experiment} & \textbf{ \boldmath $\delta m$ (GeV)}  & \textbf{ \boldmath $\tau_{\mathcal{B - \mathcal{\bar B}}}$ (s)}\\ 
\hline
\hline
\multirow{2}{*}{$n$} & ILL Grenoble (free $n$) & $7.7 \times 10^{-33}$ & $0.86 \times 10^8$~\cite{Baldo-Ceolin:1994hzw} \\[0.1em]
& Super-K (bound $n$)& $1.4 \times 10^{-33}$ & $4.7 \times 10^8$~\cite{Super-Kamiokande:2020bov} \\ [0.1em] \hline 
$\Lambda$ & BESIII (``quasi-free'' limit) & $2.1 \times 10^{-18}$ & $1.4 \times 10^{-7}$~\cite{BESIII:2024gcd} \\ [0.1em]
\hline
\end{tabular}
\caption{Current experimental limits on $\mathcal{B}-\bar{\mathcal{B}}$ oscillations for $\mathcal{B}=n, \, \Lambda$. In the second column, we indicate the relevant experiment and baryon configuration. In the third column, we show the upper limit on $\delta m$, whereas in the fourth column we show the lower bound on $\tau_{\mathcal{B - \mathcal{\bar B}}}$. See Refs.~\cite{Phillips:2014fgb,Aitken:2017wie} for more information on the experimental bounds on $|\Delta B|=2$ hadronic transitions.}
\label{tab:experimentalbounds} 
\end{table}

In Table~\ref{tab:experimentalbounds} we summarise all experimental bounds on $n - \bar n$ and $\Lambda - \bar \Lambda$ oscillations. Although the most stringent bounds on $\delta m_n$ currently arise from the Super-K search for bound neutron oscillations~\cite{Super-Kamiokande:2020bov}, the direct search for free $n - \bar n$ oscillations at the ILL Grenoble experiment sets a lower limit of $\tau_{n - \bar n} > 0.86 \times 10^8$ s~\cite{Baldo-Ceolin:1994hzw}. On the other hand, for hyperon transitions, the extremely short lifetime of the $\Lambda$ baryon precludes its interactions with the matter inside the experimental detectors of BESIII. As a result, studies of $\Lambda-\bar \Lambda $ oscillations are limited to the experiments in the ``quasi-free'' limit in the presence of a weak external magnetic field  (1 T in Tab.~\ref{tab:experimentalbounds}).

\subsection{\texorpdfstring{$|\Delta B|=2$}{Delta B =2} EFT Theoretical Framework} \label{sec:eftframework}

We now turn to the microscopic structure of the quark-level operators responsible for $|\Delta B| = 2$ hadronic transitions. The $|\Delta B| = 2$ effective interactions discussed in Sec.~\ref{sec:HeffB2} must originate from some extension of the SM, operative at an as-yet unknown scale denoted by $\Lambda_{\rm BNV}$. The off-diagonal elements $\delta m$ in the effective Hamiltonian $\mathcal{H}_{\rm eff}$ of Eq.~\eqref{eq:HeffB2} are generated by local effective operators that have been extensively studied in the context of $n - \bar n$ oscillations~\cite{Rao:1983sd,Chang:1980ey,Kuo:1980ew}. Such hadronic transitions involve six quarks in the low-energy effective field theory (LEFT), which is invariant under $\mathrm{SU(3)}_{\rm QCD} \times \mathrm{U(1)}_{\rm EM}$. These LEFT operators can arise either from the three dimension-9 $|\Delta B|=2$ operators in the SMEFT, or from higher-dimensional ($d \geq 11$) $|\Delta B|=2$ SMEFT operators, as discussed in Appendix~\ref{sec:4scalars}.\footnote{Note that operators of higher mass dimension are increasingly suppressed by powers of the UV scale. Consequently, minimal tree-level UV completions typically focus on models that generate the dimension-9 operators~\cite{Arnold:2012sd,Gardner:2018azu}.}

At present, experimental searches for baryon oscillations are limited to a few neutral baryons, most notably, the neutron ($n$), the Lambda baryon ($\Lambda$), and the b-flavored xi ($\Xi_b$). Here, we focus on the $|\Delta B| = 2$ six-quark operators corresponding to $\Delta S = 0$ and $\Delta S = 2$ transitions. In general, the off-diagonal elements $\delta m$ in Eq.~\eqref{eq:HeffB2} arise from the effective Lagrangian at the quark level,
\begin{equation} \label{eqLagrangianB2}
\mathcal{L}_{\rm eff}^{|\Delta B|=2} = \sum_i \mathcal{C}_i\,\mathcal{O}_i \; ,
\end{equation}
where the relevant six-quark operators take the schematic forms $\mathcal{O}_i \sim (udd)^2$ or $\mathcal{O}_i \sim (uds)^2$, depending on whether one is considering $n-\bar n$ or $\Lambda-\bar \Lambda$ oscillations. We leave the quark chiralities, Lorentz, and gauge indices unspecified, and the quark fields are expressed in the mass basis. The corresponding Wilson coefficients (WCs) can be written as
\begin{equation}
\mathcal{C}_i \sim \frac{c_i}{\Lambda_{\rm BNV}^5} \, ,
\end{equation}
where $\Lambda_{\rm BNV}$ denotes the UV scale associated with the BNV phenomena, and $c_i$ are dimensionless coefficients expected to be at most of $\mathcal{O}(1)$. The hadronic matrix elements relate to the WCs through
\begin{equation} \label{eq:defdeltam}
\delta m_{\mathcal{B}} = \braket{\bar{\mathcal{B}} | \mathcal{H}_{\rm eff} | \mathcal{B}} = \sum_i \mathcal{C}_i \braket{\bar{\mathcal{B}} | \mathcal{O}_i | \mathcal{B}} \, ,
\end{equation}
where $\mathcal{B} = n$ or $\Lambda$. Explicitly, we define
\begin{align}
\delta m_n &\equiv \braket{\bar n | \mathcal{H}_{\rm eff} | n } \, ,  \\
\delta m_\Lambda &\equiv \braket{\bar \Lambda | \mathcal{H}_{\rm eff} | \Lambda } \, .
\end{align}

The matrix elements $\braket{\bar{\mathcal{B}} | \mathcal{O}_i | \mathcal{B}}$ are purely hadronic quantities that must be computed using lattice QCD techniques. For $n-\bar n$ oscillations, the matrix elements have been evaluated on the lattice and found to be of order $\braket{\bar n | \mathcal{O}_i | n} \simeq 10^{-5}~\mathrm{GeV}^6$~\cite{Rinaldi:2018osy,Rinaldi:2019thf}. However, the analogous computation for $\Lambda-\bar \Lambda$ oscillations is not yet available. In the absence of such results, one may use the neutron value as an estimate, consistent up to $\mathcal{O}(1)$ factors with $\Lambda_{\rm QCD}^6$, where $\Lambda_{\rm QCD} \sim 200~\mathrm{MeV}$. Hence, a reasonable approximation is
\begin{equation}
\braket{\bar \Lambda | \mathcal{O}_i | \Lambda} \simeq \braket{\bar n | \mathcal{O}_i | n} + \mathcal{O}\left(\frac{m_{u,d}}{m_s}\right) \, ,
\end{equation}
where $\mathcal{O}\left(\frac{m_{u,d}}{m_s}\right)$ parametrises the breaking of the approximate flavor $\rm{SU(3)}$ symmetry. This estimate will be used throughout this work for all subsequent numerical analyses.

\subsubsection*{Dimension-9 \texorpdfstring{$|\Delta B|=2$}{deltab2} LEFT operators}

The case of $ n - \bar n $ oscillations has been extensively studied over the past decades, and the corresponding EFT framework is well established in terms of LEFT and SMEFT operators~\cite{Chang:1980ey,Kuo:1980ew,Rao:1982gt,Rao:1983sd,Caswell:1982qs}. Since the generalisation to $\Lambda - \bar \Lambda$ oscillations is straightforward, we first review the well-known case of $n - \bar{n}$ oscillations, and then present results for the $\Lambda - \bar \Lambda$ system. 

At the level of the LEFT, one can write three possible colour-singlet and electrically-neutral operators contributing to $n - \bar n$ at the quark level
\begin{align}
    \mathcal{O}_{\chi_1 \chi_2 \chi_3}^1 &= (\bar u^c_i P_{\chi_1} u_j)(\bar d^c_k P_{\chi_2} d_l)(\bar d^c_m P_{\chi_3} d_n)T^{SSS}_{\{ij\}\{kl\}\{mn\}} \, , \label{eq:LEFTop1}\\
    \mathcal{O}_{\chi_1 \chi_2 \chi_3}^2 &= (\bar u^c_i P_{\chi_1} d_j)(\bar u^c_k P_{\chi_2} d_l)(\bar d^c_m P_{\chi_3} d_n)T^{SSS}_{\{ij\}\{kl\}\{mn\}} \, ,\label{eq:LEFTop2}\\
    \mathcal{O}_{\chi_1 \chi_2 \chi_3}^3 &= (\bar u^c_i P_{\chi_1} d_j)(\bar u^c_k P_{\chi_2} d_l)(\bar d^c_m P_{\chi_3} d_n)T^{AAS}_{[ij][kl]\{mn\}} \, , \label{eq:LEFTop3}
\end{align}
where parentheses denote Lorentz contractions, and  $\chi_i = P, L$ indicates the chirality of the quark bilinear with $P_{\chi_i}$ representing the corresponding chiral projection. The last set of indices $i,j,k,l,m,n=1,2,3$ are $\rm SU(3)_C$ colour indices, which are contracted into the colour structures $T^{XYZ}$, characterized by definite symmetry properties under the exchange of colour index pairs, specified by the superscripts $X, Y, Z$, which can be $S$ or $A$ to denote a symmetric or antisymmetric contraction, respectively. Explicitly, they are defined as~\cite{Rao:1982gt,Kuo:1980ew}
\begin{align}
    T^{SSS}_{\{ij\}\{kl\}\{mn\}} &= \epsilon_{ikm} \epsilon_{jln} + \epsilon_{jkm} \epsilon_{iln} + \epsilon_{ilm} \epsilon_{jkn} + \epsilon_{ikn} \epsilon_{jlm} \, , \label{eq:colorstructure1}\\
    T^{AAS}_{[ij][kl]\{mn\}} &= \epsilon_{ijm} \epsilon_{kln} + \epsilon_{ijn} \epsilon_{klm} \, . \label{eq:colorstructure2}
\end{align}
Note that, at the level of the LEFT, since each $\chi_I$ may take two values, each of the three operators $\mathcal{O}^I$ written in Eqs.~\eqref{eq:LEFTop1}--\eqref{eq:LEFTop3} is made out of eight distinct chiral combinations, naively this would lead to 24 distinct operators. Nevertheless, there exist additional relations given by $\mathcal{O}_{\chi_1 RL}^1 = \mathcal{O}_{\chi_1 LR}^1$, and $\mathcal{O}_{RL \chi_3}^{2,3} = \mathcal{O}_{LR \chi_3}^{2,3}$, that reduce by $2\times3=6$ the number of independent combinations, leading to 18 independent ones in total~\cite{Rao:1982gt,Kuo:1980ew}. Finally, the relation $\mathcal{O}_{\chi \chi \chi'}^2- \mathcal{O}_{\chi \chi \chi'}^1 = 3 \mathcal{O}_{\chi \chi \chi'}^3$, for two distinct chiralities $\chi$, and $\chi'$, pointed out in Ref.~\cite{Caswell:1982qs} further reduces the number of independent operators by four, leading to only 14. These operators are not independent under isospin symmetry transformations.

For $ \Lambda - \bar \Lambda $ oscillations, one can repeat the previous analysis, keeping in mind the presence of strange and anti-strange quarks. The complete operator basis can be obtained following the early work of Ref.~\cite{Rao:1983sd}, where the analysis in terms of general flavour indices was developed. In our case, with two strange quarks, the presence of an additional quark flavour increases the number of independent operators from 14 to 52, as verified using the \textsc{Mathematica} package \texttt{Sym2Int}~\cite{Fonseca:2017lem,Fonseca:2019yya}. From a phenomenological perspective, however, it is more interesting to identify which EFT operators induce exclusively $ \Lambda - \bar \Lambda $ oscillations without simultaneously generating $ n - \bar n $ transitions. Whenever a left-handed strange quark is involved, one will also have an analogous operator with the replacement $ s_L \to d_L $ through fermion mixing, which is CKM suppressed (for more discussion, see \cite{Beneke:2024hox}). Since the experimental bounds on $ n - \bar n $ oscillations are significantly more stringent than those on $ \Lambda - \bar \Lambda $ as it can be seen in Tab.~\ref{tab:experimentalbounds}, such EFT contributions are not expected to be phenomenologically relevant. On the other hand, the presence of right-handed strange quarks $ s_R $ renders the two baryon oscillations disconnected at tree level. In this case, a transition $ s_R \to d_R $ would require a mechanism such as $W$-boson exchange within the SM, leading to additional suppression of the operators contributing to $ n - \bar n $ oscillations. Further discussion of these aspects can be found in Secs.~\ref{sec:exoticpheno} and~\ref{sec:uvmodel}, and a comprehensive derivation of the full operator basis will be presented in a forthcoming publication.

\subsubsection*{Dimension-9 \texorpdfstring{$|\Delta B|=2$}{Delta B =2} SMEFT operators}

From a top-down point of view, out of all the possibilities at the level of the LEFT, only a few of these operators can be obtained as a low-energy realisation of a full SM gauge group $G_{\rm SM} \equiv \rm{SU}(3)_C \times \rm{SU}(2)_L \times \rm{U}(1)_Y$ invariant dimension-9 $|\Delta B|=2$ SMEFT operator from SMEFT-LEFT tree-level matching.\footnote{To easily visualize this note that $\mathcal{O}_{LLL}^I$ for any $I=1,2,3$ can only arise from an SMEFT operator involving six $\rm SU(2)_L$ doublets $Q$, requiring an additional $H^\dagger H^\dagger$ insertion to account for $\rm U(1)_Y$ invariance. This corresponds to the simplest model with a quartic $|\Delta B|=2$ potential explained in Eq.~\eqref{eq:app1} of App.~\ref{sec:4scalars}.} Since our assumption is that the new exotic scalars incorporated into the SM particle content to account for $|\Delta B|=2$ processes have masses above the electroweak scale, as will become clear in the next section, a more appropriate framework to study this phenomenon is the SMEFT. There exist three dimension-9 $|\Delta B|=2$ operators with distinct particle content that can generate $n - \bar n$ at tree level. These operators carry an odd charge under $\Delta (B-L)/2$, ensuring they always appear at odd operator dimensions~\cite{Kobach:2016ami,deGouvea:2014lva}. Consequently, $n - \bar n$ also arises at dimensions 11, 13, and beyond, with each successive contribution increasingly suppressed by higher powers of the UV scale $\Lambda$. We highlight some interesting phenomenology in the Appendix~\ref{sec:4scalars}, where quartic couplings associated with scalar UV completions of dimension-11 operators are introduced.

Similarly, at the level of the SMEFT, there are three dimension-9 $|\Delta B|=2$ operators with distinct SM field content that contribute to $|\Delta B|=2$ processes. Two of these operators, however, can each be written as two independent terms, so that one can define a total of five operators once all gauge and Lorentz indices are contracted. We write these operators for $n_g = 3$ fermion generations as
\begin{equation}
    \mathcal{L} \supset \sum_{i}^3\frac{1}{\Lambda_{\rm BNV}^5}[\mathcal{C}_i]_{prstuv} [\mathcal{O}_i]_{prstuv} \; ,
\end{equation}
where~\cite{Liao:2020jmn,Li:2020xlh}
\begin{align}
    [\mathcal{O}_1]_{prstuv} &= \epsilon_{ijk}\epsilon_{lmn}(\bar u_{R\,p}^{c\, i}d_{R\,r}^{l})( \bar u_{R\,s}^{c \, j} d_{R\,t}^{m})(\bar d_{R\,u}^{c \, k} d_{R\,v}^{n})\label{eq:op1of5} \, , \\
    [\mathcal{O}_2]_{prstuv} &= \epsilon_{imk}\epsilon_{ljn}(\bar u_{R\,p}^{c\, i}d_{R\,r}^{l})( \bar u_{R\,s}^{c \, j} d_{R\,t}^{m})(\bar d_{R\,u}^{c \, k} d_{R\,v}^{n})\label{eq:op2of5}  \, , \\
    [\mathcal{O}_3]_{prstuv} &= \epsilon_{ijk}\epsilon_{lmn}\epsilon_{IJ}(\bar u_{R\,p}^{c\, i}d_{R\,r}^{l})( \bar d_{R\,s}^{c \, j} d_{R\,t}^{m})(\bar Q_{L\,u}^{c \, k \, I} Q_{L\,v}^{n \, J})\label{eq:op3of5}  \, , \\
    [\mathcal{O}_4]_{prstuv} &= \epsilon_{ijk}\epsilon_{lmn}\epsilon_{IK}\epsilon_{JL}(\bar d_{R\,p}^{c\, i}d_{R\,r}^{l})( \bar Q_{L\,s}^{c \, j \, I} Q_{L\,t}^{m \, J})(\bar Q_{L\,u}^{c \, k \, K} Q_{L\,v}^{n \, L})\label{eq:op4of5}  \, , \\
    [\mathcal{O}_5]_{prstuv} &= \epsilon_{ijn}\epsilon_{lmk}\epsilon_{IK}\epsilon_{JL}(\bar d_{R\,p}^{c\, i}d_{R\,r}^{l})( \bar Q_{L\,s}^{c \, j \, I} Q_{L\,t}^{m \, J})(\bar Q_{L\,u}^{c \, k \, K} Q_{L\,v}^{n \, L})\label{eq:op5of5}  \, . 
\end{align}
Here, $i,j,k,l,m,n$ denote $\rm{SU(3)_C}$ indices, $I,J,K,L$ denote $\rm{SU(2)_L}$ indices, and $p,r,s,t,u,v$ denote flavour indices. The chiral quarks transform under $G_{\rm SM}$ as
\begin{equation}
    Q_L \sim (\textbf{3}, \textbf{2},1/6) \,,
    \hspace{1cm}
    u_R \sim (\textbf{3}, \textbf{1},2/3) \, ,
    \hspace{1cm}
    d_R \sim (\textbf{3}, \textbf{1},-1/3) \, . \nonumber
\end{equation}
If we restrict to first-generation quarks only, one obtains the four well-known SMEFT operators contributing to $n - \bar n$ oscillations at tree level, given by~\cite{Rao:1983sd,Rao:1982gt}
\begin{align}
    \mathcal{O}_1 &= T^{SSS}_{\{ij\}\{kl\}\{mn\}}(\bar u^{c\, i}_R u^{j}_R)( \bar d^{c \, k}_R d^{l}_R)(\bar d^{c \, m}_R d^{n}_R)\label{eq:b2smeftopsRRRnew} \, , \\
    \mathcal{O}_2 &= T^{SSS}_{\{ij\}\{kl\}\{mn\}}(\bar u^{c\, i}_R d^{j}_R)( \bar u^{c \, k}_R d^{l}_R)(\bar d^{c \, m}_R d^{n}_R)\label{eq:b2smeftopsRRRnew1}  \, , \\
    \mathcal{O}_3 &= T^{AAS}_{[ij][kl]\{mn\}}\epsilon_{IJ}(\bar Q^{c \, iI}_L Q^{jJ}_L)(\bar u^{c \, k}_R d^l_R)(\bar d^{c \, m}_R d^n_R)\label{eq:b2smeftopsLRRnew}  \, , \\
    \mathcal{O}_4 &= T^{AAS}_{[ij][kl]\{mn\}}\epsilon_{IK}\epsilon_{JL}(\bar Q^{c \, i\, I}_L Q^{j \, J}_L)(\bar Q^{c \, k \, K}_L Q^{l \, L}_L)(\bar d^{c \, m}_R d^n_R)\label{eq:b2smeftopsLLRnew}  \, .
\end{align}

For $ \Lambda - \bar \Lambda $ oscillations, the complete operator basis can be obtained following the early work of Ref.~\cite{Rao:1983sd}, as for the LEFT scenario. In our case, with two strange quarks, the presence of an additional quark flavour increases the number of independent operators from 4 to 20, as verified using the \textsc{Mathematica} package \texttt{Sym2Int}~\cite{Fonseca:2017lem,Fonseca:2019yya}.

\subsection{Scalar mediated \texorpdfstring{$|\Delta B|=2$}{Delta B =2} processes} \label{sec:scalarmediated}

We consider UV completions of these operators at tree level via exotic scalar fields that have masses above the EW scale. In this work, we study the $|\Delta B|=2$ processes mediated only via scalar leptoquark couplings, setting aside completions involving massive vector bosons and fermions. For completeness, Appendix~\ref{sec:fermionscalartopology} presents an exhaustive list of the UV completions involving two scalars and one fermion. We choose not to explore $|\Delta B|=2$ topologies involving vector LQs, as we remain agnostic about the structure of the underlying UV theory and the spontaneous symmetry breaking mechanism responsible for their mass generation. In Sec.~\ref{sec:3scalars} we analyze scalar UV completions via trilinear interactions in the scalar potential for the operators given in Eqs.~\eqref{eq:op1of5}--\eqref{eq:op5of5}, while in Appendix~\ref{sec:4scalars} we investigate completions involving quartic scalar interactions for $|\Delta B|=2$ processes.

\subsubsection{UV completions with trilinear couplings} \label{sec:3scalars}

\begin{figure}[ht]
    \centering
    \includegraphics[width = 5.5cm]{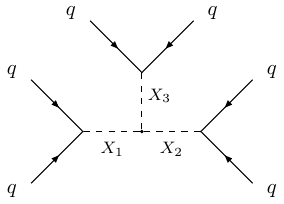}
    \caption{Triple scalar interaction leading to $|\Delta B|=2$ processes at tree-level. The label $q$ denotes generic quark-type fermions. See main text for further details.}
    \label{fig:scalarUVtopology}
\end{figure}

In this section, we focus on all UV scalar multiplets that couple linearly (at the renormalisable level) to a quark bilinear, i.e. they have diquark couplings. These trilinear couplings were originally introduced in Ref.~\cite{Mohapatra:1980qe}, and have been studied later on mainly in the context of baryogenesis in Refs.~\cite{Baldes:2011mh,Herrmann:2014fha,Babu:2012vc}. The topology leading to $|\Delta B|=2$ processes is depicted in Figure~\ref{fig:scalarUVtopology}.

These scenarios, referred to as {\it Minimal} models in Refs.~\cite{Arnold:2012sd,Gardner:2018azu}, may also include scalar multiplets with leptoquark couplings, such as $S_1$ and $S_3$, which mediate proton decay at tree level. In the absence of additional UV symmetries forbidding $|\Delta B| = 1$ processes, their masses are typically constrained to lie around $10^{15\text{--}16}~\text{GeV}$~\cite{Fajfer:2023gfi}. The quantum numbers of these scalars and their diquark interactions relevant for $|\Delta B| = 2$ transitions are summarised in Table~\ref{tab:definitionsLQ}, and the list of all possible tree-level UV completion of the dimension-9 $|\Delta B|=2$ SMEFT operators in this specific topology involving 2 or 3 new exotic scalars $X_i$, i.e. $\mu X_1X_1X_2$ and $\mu X_1 X_2 X_3$,\footnote{The latter possibility was not explored in Ref.~\cite{Arnold:2012sd}.} can be found in Table~\ref{tab:DeltaB2models}.

\begin{table}[ht]

\centering
\renewcommand{\arraystretch}{1.5}
\setlength{\tabcolsep}{14pt}

\begin{tabular}{|c|c|}
\hline
\textbf{Scalar} & \textbf{Diquark Interaction}\\ 
\hline
\hline
$S_1\sim(\mathbf{ \bar 3},\mathbf{1},1/3) $ & $ (y^{L}_{S_1} )_{\{ij\}} \,\bar{Q}^{c\,i}_L S_1^* Q^j_L  +  (y^{R}_{S_1} )_{ij}\, \bar u_R^{c\, i} S_1^* d_{R}^{j} +\mathrm{h.c.}$ \\[0.1em]
 
$S_3 \sim(\mathbf{ \bar 3},\mathbf{3},1/3) $ & $ (y^{L}_{S_3} )_{[ij]} \,\bar{Q}^{c\,i}_L (\vec \tau \cdot \vec S_3)^* Q^j_L  +\mathrm{h.c.}$ \\[0.1em]

$\tilde{S}_1 \sim (\mathbf{ \bar 3},\mathbf{1},4/3) $ & $  (y^{R}_{\tilde{S}_1} )_{[ij]}\, \bar u_{R}^{c\,i} \tilde S_1^* u_{R}^{j} +\mathrm{h.c.}$ \\ [0.1em]

$\bar{S}_1 \sim (\mathbf{ \bar 3},\mathbf{1}, -2/3) $ & $ (y^{R}_{\bar{S}_1} )_{[ij]}\, \bar d_{R}^{c\, i} \bar S_1^* d_{R}^{j} +\mathrm{h.c.}$ \\[0.1em] \cdashline{1-2}

$\Omega_1 \sim (\mathbf{ \bar 6},\mathbf{1}, -1/3) $ & $ (y^{L}_{\Omega_1} )_{[ij]} \,\bar{Q}_L^{c\,i} \Omega_1 Q^j_L  +  (y^{R}_{\Omega_1} )_{ij}\, \bar u_{R}^{c\, i} \Omega_1 d_{R}^{j} +\mathrm{h.c.}$ \\[0.1em]
 
$\Upsilon \sim(\mathbf{ \bar 6},\mathbf{3}, -1/3) $ & $ (y^{L}_{\Upsilon} )_{\{ij\}} \,\bar{Q}^{c\,i}_L (\vec \tau \cdot \vec \Upsilon) Q^j_L  +\mathrm{h.c.}$ \\[0.1em]

$\Omega_2 \sim (\mathbf{ \bar 6},\mathbf{1},2/3) $ & $  (y^{R}_{\Omega_2} )_{\{ij\}}\, \bar d_{R}^{c\, i} \Omega_2 d_{R}^{j} +\mathrm{h.c.}$ \\ [0.1em]

$\Omega_4 \sim (\mathbf{ \bar 6},\mathbf{1}, -4/3) $ & $ (y^{R}_{\Omega_4} )_{\{ij\}}\, \bar u_{R}^{c\, i} \Omega_4 u_{R}^{j} +\mathrm{h.c.}$ \\

\hline 
\end{tabular}
\caption{Scalars and their representations under $G_{\rm SM}$, together with their diquark interactions with SM quark bilinears. The first three multiplets also admit quark–lepton couplings, which are not shown here. The dashed line separates the multiplets according to their colour representation, i.e.~anti-triplets and anti-sextets. Explicit $\mathrm{SU(3)_C}$ and $\mathrm{SU(2)_L}$ indices are omitted. Flavour indices are denoted by $i,j$, with curly (square) brackets indicating symmetry (antisymmetry) under interchange. See Ref.~\cite{Dorsner:2016wpm} for further details.}
\label{tab:definitionsLQ} 
\end{table}

\begin{table}[ht]
\centering
\renewcommand{\arraystretch}{1.5}
\setlength{\tabcolsep}{14pt}

\begin{tabular}{|c|c|c|c|}
\hline
\textbf{Interaction} & \textbf{Order \boldmath $\delta m_n$} & \textbf{Order \boldmath $\delta m_\Lambda$} & \textbf{Proton Decay} \\ 
\hline
\hline
$S_1S_1\Omega_2^*$ & Tree-level & Tree-level & \cmark \\
$S_3S_3\Omega_2^*$ & Tree-level & Tree-level & \cmark \\
$\bar S_1\bar S_1\Omega_4^*$ & 2-loops & Tree-level & \xmark \\
$\Omega_1\Omega_1\Omega_2$ & Tree-level & Tree-level & \xmark \\
$\Upsilon\Upsilon\Omega_2$ & Tree-level & Tree-level & \xmark \\
$\Omega_2\Omega_2\Omega_4$ & Tree-level & Tree-level & \xmark \\ 
\cdashline{1-4}
$S_1\bar S_1\Omega_1^*$ & 1-loop & Tree-level & \cmark \\
$S_3\bar S_1\Upsilon^*$ & 1-loop & Tree-level & \cmark \\
$\bar S_1\tilde S_1\Omega_2^*$ & 1-loop & 1-loop & \cmark \\
\hline
\end{tabular}

\caption{%
Schematic trilinear couplings of the types $\mu X_1X_1X_2$ and $\mu X_1X_2X_3$ 
arising from two or three exotic scalar multiplets of Tab.~\ref{tab:definitionsLQ}, 
leading to $|\Delta B| = 2$ hadronic transitions via the diagram of Fig.~\ref{fig:scalarUVtopology}.
The dashed line separates the two coupling types, $\mu X_1X_1X_2$ and $\mu X_1X_2X_3$. Each trilinear interaction is implicitly accompanied by a dimensionful coupling $\mu$. Models were cross-checked using the \textsc{Mathematica} package \texttt{Sym2Int}~\cite{Fonseca:2017lem,Fonseca:2019yya}. The second (third) column shows the loop order generating $\delta m_n$ ($\delta m_\Lambda$). The last column marks models that also induce proton decay.}
\label{tab:DeltaB2models}
\end{table}

For the first class of models, characterised by the coupling $\mu X_1X_1X_2$, we have identified two new models involving the scalar LQs $S_{1,3}$, that mediate $n-\bar{n}$ oscillations at tree level. This expands the set of previously well-known scenarios of Ref.~\cite{Arnold:2012sd}. In contrast, the second class of models, characterised by the coupling $\mu X_1X_2X_3$, has been largely neglected. This is due to two key reasons: the inevitable presence of a proton decay-mediating scalar LQ and a structural suppression of tree-level $n-\bar{n}$ oscillations, the latter arising from the antisymmetric couplings of the colour anti-triplets $\bar{S}_1$ and $\tilde{S}_1$, as indicated in Tab.~\ref{tab:DeltaB2models}.

The scalar LQs $S_1$, $S_3$, and $\tilde{S}_1$ can mediate proton decay, but this could be forbidden by imposing $L$ conservation, which would prevent the simultaneous presence of a lepton-quark coupling and diquark coupling. Such a conservation law would potentially relax the stringent mass bounds on these scalars. However, imposing exact $L$ conservation is theoretically unmotivated. The observed phenomenon of neutrino oscillations demonstrates that lepton flavor is violated, and if neutrinos are Majorana particles, $L$ is also broken. Assuming $L$ is broken at a high scale, as is $B$ in these models, is therefore more natural. Consequently, the proton decay constraints cannot be evaded, requiring the masses of these LQs to satisfy $M \gtrsim 10^{15}$ GeV, in accordance with current limits~\cite{Fajfer:2023gfi,Dorsner:2016wpm}.

Given these stringent bounds, it follows that $n-\bar n$ oscillations mediated by such heavy scalars are phenomenologically negligible. Assuming for simplicity the couplings in the UV theory to be $\mathcal{O}(1)$, after integrating out at tree level the scalar LQs $S_{1,3}$ and the additional BSM scalar $\Omega_2$ yields a WC for the $|\Delta B|=2$ operator of the Lagrangian in Eq.~\eqref{eqLagrangianB2} of the order
\begin{equation}
     \mathcal{C}_i \sim \frac{\mu}{M_{S_i}^4 M_{\Omega_2}^2}\, , \label{eq:d9WCnaive} 
\end{equation}
where $M_{S_i}$ and $M_{\Omega_2}$ stand for the masses of the scalar LQ $S_{i}$, and $\Omega_2$, respectively. Using the lower bounds on $M_{S_i}$ derived from the non-observation of proton decay, together with the lower limit on the $\Omega_2$ mass, set around the TeV scale by direct searches~\cite{CMS:2019gwf}, and taking the limiting case $\mu \sim M_{S_i}$, we estimate the Wilson coefficient governing $n-\bar n$ oscillations to be of order $\mathcal{C}_i \sim 10^{-53}\,\text{GeV}^{-5}$. This value lies many orders of magnitude below the current experimental sensitivity, as inferred from Super-Kamiokande limits and summarised in Tab.~\ref{tab:experimentalbounds}, where one finds $\mathcal{C}_i \lesssim 10^{-28}$ GeV$^{-5}$ after using Eq.~\eqref{eq:defdeltam} with
the numerical nuclear matrix elements $\braket{\bar n| \mathcal{O}_i|n} \sim 10^{-5} \; \rm GeV^6$~\cite{Rinaldi:2018osy,Rinaldi:2019thf}. Therefore, we conclude that under our assumptions, $n - \bar n$ induced by these two models is far from being tested in future experiments and we dismiss them in the rest of this work.

On the other hand, for the other three models of Tab.~\ref{tab:DeltaB2models} that induce $n - \bar n$ at tree-level, which correspond to Models 1, 2, and 3 of Ref.~\cite{Arnold:2012sd}, one can use the experimental bounds of Tab.~\ref{tab:experimentalbounds} along with the EFT framework developed in Sec.~\ref{sec:eftframework} to find
\begin{equation}
    \frac{10^{-33} \; \rm GeV  }{\delta m_n^{\rm exp}} \lesssim \;\left(\frac{\Lambda_{\rm BNV}}{300 \; \rm TeV}\right)^5  \; , \label{eq:roughestimatennbar}
\end{equation}
where we have neglected $\mathcal{O}(1)$ factors, and we have defined an 
effective UV scale $\Lambda_{\rm BNV}$ as
\begin{equation}
    \Lambda_{\rm BNV}^{-5} \equiv \frac{\mu}{M_S^4M_\Omega^2} \; .
\end{equation}
Therefore, in these models, experimental constraints on $n - \bar n$ oscillations put a stringent lower bound on the effective UV scale $\Lambda_{\rm BNV}$ up to $10^2-10^3$ TeV, thus unreachable in foreseeable collider experiments.

\section{Exotic \texorpdfstring{\boldmath $|\Delta B|=2$}{Delta B =2} phenomenology} \label{sec:exoticpheno}

Recent results reported by BESIII collaboration \cite{BESIII:2024gcd} have yielded direct constraints on the $\Lambda-\bar \Lambda$ oscillations. It would be advantageous to see if indirect constraints on the oscillation parameters can be derived, similar to those obtained for the $|\Delta B| = 1$ decays~\cite{Beneke:2024hox,Crivellin:2023ter,Hou:2005iu}. In this section, we employ the chiral Lagrangian to derive indirect constraints on $\delta m_\Lambda$ from two observables: the limits on $\delta m_n$ obtained in $n-\bar n$ oscillation searches, and the current bounds on $p \, p \to K^+ K^+$ from Super-Kamiokande.

\subsection{Constraints on \texorpdfstring{$\delta m_\Lambda$}{deltamlambda} from \texorpdfstring{$n-\bar n$}{nnb} oscillations}
\label{sec:constraintsoscillations}

In Ref.~\cite{Luk:2007LambdaOscillation}, it was argued that the existence of $n - \bar n$ oscillations would necessarily imply the possibility of $\Lambda- \bar \Lambda$ oscillations. The proposed mechanism relies on the weak decay $\Lambda \to n\pi$, with the pionic loop and the $n-\bar n$ transition jointly inducing $\Lambda-\bar\Lambda$ mixing. This dynamics results in a strong suppression of order $G_F^2$, where $G_F = 1.16 \times 10^{-5}\,\text{GeV}^{-2}$ denotes the Fermi constant. Conversely, there is no reason to forbid operators containing two $s$ quarks instead of two $d$ quarks. The opposite can be true: a UV theory could exist that does not allow the presence of all four $d$ quarks in the $|\Delta B|=2$ operator, for instance, in models involving $\bar S_1$. 

So far, we have discussed the $|\Delta B|=2 $ transition on the quark level. However, these quark-level amplitudes must be matched onto the baryonic and mesonic effective theory at the hadronic scale, $\mu \sim m_n \simeq 1~\text{GeV}$. To this end, we first write down the effective operator describing the $n - \bar n$ transition~\cite{Berezhiani:2015uya,Berezhiani:2018xsx}
\begin{equation}
\mathcal{L}^{|\Delta B|=2}(n -\bar n) = -\frac{1}{2} \, \delta m_n\left[ n^T C n + \bar{n} C \bar{n}^T \right],
\end{equation}
and also for the $\Lambda -\bar \Lambda$ transition 
\begin{equation}
\mathcal{L}^{|\Delta B|=2}(\Lambda -\bar \Lambda) = -\frac{1}{2} \, \delta m_\Lambda \left[ \Lambda^T C \Lambda+ \bar \Lambda C \bar\Lambda^T \right]. \label{eq:deltamLambda}
\end{equation}
The $\delta m_i$ have a dimension of mass, $C$ stands for the charge conjugation, and in our notation it is defined as $C= i \, \gamma^0 \gamma^2$. This results from the $|\Delta B|=2 $ transition $\mathcal{L} = \delta m \, \bar \Psi^c \Psi$, where $\bar \Psi^c= \Psi^T C$. To constrain $\delta m_\Lambda$ from $\delta m_n$, one can make use of Baryon Chiral Perturbation Theory ($B\chi PT$), whose Lagrangian describes the interactions between baryons and mesons, and can be written in terms of the meson $\phi$ and baryon $B$ fields \cite{JLQCD:1999dld,Claudson:1981gh}

\begin{eqnarray}
\phi&=&\left(\begin{array}{ccc}
     \frac{1}{\sqrt{2}}\pi^0+\frac{1}{\sqrt{6}}\eta &\pi^+&K^+ \\
     \pi^-&-\frac{1}{\sqrt{2}}\pi^0+\frac{1}{\sqrt{6}}\eta &K^0 \\
     K^-&{\bar K}^0&-\frac{2}{\sqrt{6}}\eta \\
             \end{array} \right), \\
B&=&\left(\begin{array}{ccc}
     \frac{1}{\sqrt{2}}\Sigma^0+\frac{1}{\sqrt{6}}\Lambda^0&\Sigma^+&p \\
     \Sigma^-&-\frac{1}{\sqrt{2}}\Sigma^0+\frac{1}{\sqrt{6}}\Lambda^0&n \\
     \Xi^-&\Xi^0&-\frac{2}{\sqrt{6}}\Lambda^0 \\
          \end{array} \right).
\end{eqnarray}

We also define the $3\times3$ special unitary matrices in terms of $\phi$

\begin{equation}
\Sigma={\rm exp}\left(\frac{2i\phi}{f}\right),\quad \xi={\rm exp}\left(\frac{i\phi}{f}\right),
\end{equation}

where $f=130.4$ MeV is the pion decay constant~\cite{ParticleDataGroup:2024cfk}. The meson and baryon fields have the following transformation properties under the flavour group $\rm SU(3)_L\times SU(3)_R$  
\begin{equation}
\Sigma\rightarrow L\Sigma R^\dagger,\quad
B\rightarrow UBU^\dagger.
\end{equation}

The lowest order of the $\rm SU(3)_L\times SU(3)_R$ chiral Lagrangian in Euclidean spacetime is given by
\begin{eqnarray}
&{\cal L}_0=&\frac{f^2}{8}\tr(\partial_\mu\Sigma)(\partial_\mu\Sigma^\dagger)
             +\tr{\bar B}(\gamma_\mu\partial_\mu+M_B)B 
         +\frac{1}{2}\tr{\bar B}\gamma_\mu[\xi\partial_\mu\xi^\dagger
             +\xi^\dagger\partial_\mu\xi]B \nonumber \\
            && +\frac{1}{2}\tr{\bar B}\gamma_\mu B[(\partial_\mu\xi)\xi^\dagger
             +(\partial_\mu\xi^\dagger)\xi] 
         -\frac{1}{2}(D-F)\tr{\bar B}\gamma_\mu\gamma_5 B
             [(\partial_\mu\xi)\xi^\dagger
             -(\partial_\mu\xi^\dagger)\xi] \nonumber \\
          && +\frac{1}{2}(D+F)\tr{\bar B}\gamma_\mu\gamma_5 
             [\xi\partial_\mu\xi^\dagger
             -\xi^\dagger\partial_\mu\xi]B \, .
             \label{ChL}
\end{eqnarray}
Experimental results for the semileptonic baryon decays 
give $F=0.46$ and $ D=0.80$~\cite {ParticleDataGroup:2024cfk}.

\subsubsection*{Tree level contributions}

\begin{figure}[h]
\centering
\includegraphics[scale=1.1]{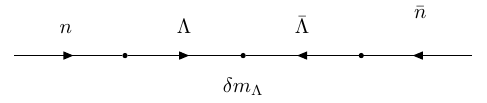}
\caption{ \label{fig:app0diagram} Extraction of $\delta m_\Lambda$ from $n- \bar n$ oscillations at tree-level. The central dot indicates the $|\Delta B| = 2$ mass insertion from Eq.~\eqref{eq:deltamLambda}. The intermediate $\Lambda$ baryons are off-shell.}
\end{figure}

Within the chiral Lagrangian approach, there is a tree tree-level diagram contributing to $\delta m_n$ which is presented in Figure~\ref{fig:app0diagram}.

To compute this tree-level diagram we need the nonleptonic weak Lagrangian. We follow  the calculations of the weak nonleptonic decays presented in Ref.~\cite{Bijnens:1985kj}, where the weak Lagrangian is given by~\cite{Ivanov:2021huf,Tandean:2002vy}
\begin{equation}
L_{\text{eff}}^{|\Delta S| = 1} = 
a\, \text{Tr} \left( \bar{B} \{ \xi^\dagger h \xi, B \} \right) 
+ b\, \text{Tr} \left( \bar{B} [ \xi^\dagger h \xi, B ] \right) 
+ \text{h.c.} \, , \label{eq:weaknonleptonic}
\end{equation}
where
\[
h = 
\begin{pmatrix}
0 & 0 & 0 \\
0 & 0 & 1 \\
0 & 0 & 0
\end{pmatrix} \, ,
\]
and $a= 0.56\, G_F\, m_\pi^2\, f =1.68\times 10^{-8}$ GeV and $b= -1.42\,G_F\,m_\pi^2\, f =-4.26\times 10^{-8}$ GeV. Expanding the Lagrangian of Eq.~\eqref{eq:weaknonleptonic} we find the term responsible for this transition
\begin{equation}
    \mathcal{L}_{eff}^{|\Delta S|=1} \supset -\frac{a+3b}{\sqrt{6}} \,\bar \Lambda\, n + \text{h.c.}\,,
\end{equation}
so that the tree-level contribution is given by
\begin{equation}
    \delta m_n \sim \left( \frac{a+3b}{\sqrt{6}}\right)^2 \frac{(m_n + m_\Lambda)^2}{(m_\Lambda^2-m_n^2)^2} \,\delta m_\Lambda \sim 6.2 \times 10^{-14} \,\delta m_\Lambda \, .
\end{equation}
Therefore, from the experimental bound on $n - \bar n$ oscillations $\delta m_n \leq 1.4 \times 10^{-33}$ GeV~\cite{Super-Kamiokande:2020bov}, one can infer the approximate upper bound for the term in Eq.~\eqref{eq:deltamLambda}
\begin{equation}
    \delta m_\Lambda \lesssim  2.3 \times10^{-20} \; \text{GeV} \;.
\end{equation}
Although the tree-level contribution is expected to dominate over the loop-level ones, for completeness we explicitly compute the two leading 1-loop diagrams: the pionic and the kaonic loops.

\subsubsection*{Pionic loop: \texorpdfstring{$\delta m_{\Lambda }$}{deltabmlambda} from \texorpdfstring{$\delta m_{n}^{\pi}$}{deltabmlambda}}

\begin{figure}[!htp]
\centering
\includegraphics[scale=1.1]{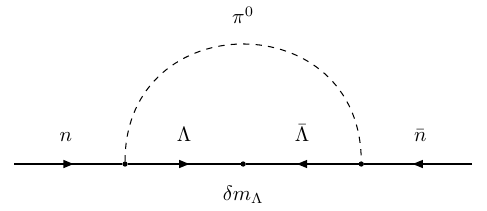}
\caption{ \label{fig:app1diagram} The extraction of $\delta m_\Lambda$ from the $n-\bar n$ oscillations through a $\pi^0$ running within the loop. The central dot indicates the $|\Delta B| = 2$ mass insertion from Eq.~\eqref{eq:deltamLambda}.
}
\end{figure}

The diagram presented in Figure~\ref{fig:app1diagram} also contributes to $\delta m_n$ at 1-loop.

In this diagram, there are two weak vertices, and to compute the corresponding amplitudes, we expand the nonleptonic weak Lagrangian of Eq.~\eqref{eq:weaknonleptonic} to first order in the meson fields. The amplitude for the $n - \bar n$ transition via the $\pi$-loop is
\begin{eqnarray}
 && \mathcal{M} (n - \bar n )^{\pi}  = {v}_n^T C  \int \,\frac{d^4 q}{(2 \pi)^4} \left( \frac{a+3 b}{2 \sqrt 3 f}\right)^2\frac{(\slashed{q} + m_\Lambda
 ) 
 \,\delta m_\Lambda \, (\slashed{q} + m_\Lambda)}{(q^2- m_\Lambda^2)^2 (q^2 - m_\pi^2) } u_n \, .
\label{Lnpi}
\end{eqnarray}
Therefore, we can write the mass mixing term induced by the $\pi$-loop as
\begin{eqnarray}
 && \delta m_n^{\pi}  =  \int \,  \frac{d^4 q}{(2 \pi)^4} \left( \frac{a+3 b}{2 \sqrt 3 f}\right)^2 \delta m_\Lambda
\frac{(q^2+ m_\Lambda^2)}{(q^2- m_\Lambda^2)^2 (q^2 - m_\pi^2) } \, .
\label{Lnpi1}
\end{eqnarray}
This integral is logarithmically divergent and can be computed in Euclidean spacetime using a hard cut-off $\Lambda$, which can be taken to be $\mathcal{O}(1)$ GeV corresponding to the scale where the validity of $B\chi PT$ breaks down. Under this prescription, and setting $\Lambda = 5$ GeV, the integral yields
\begin{equation}
    \delta m_n^\pi \simeq 4 \times  10^{-16} \,\delta m_\Lambda \, ,
\end{equation}
which translates into the approximate upper bound
\begin{equation}
    \delta m_\Lambda^\pi \lesssim 3.5 \times 10^{-18} \; \text{GeV} .
\end{equation}

\subsubsection*{Kaonic loop: \texorpdfstring{$\delta m_{\Lambda }$}{deltabmlambda} from \texorpdfstring{$\delta m_{n}^{K} $}{deltabmlambdaK}}

\begin{figure}[htp!]
\centering
\includegraphics[scale=1.1]{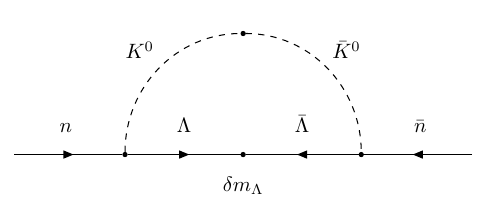}
\caption{ \label{fig:app2diagram} The extraction of $\delta m_\Lambda$ from the $n-\bar n$ oscillation through $K-\bar K$ mixing. The central dot indicates the $|\Delta B| =2$ mass insertion from Eq.~\eqref{eq:deltamLambda}.
}
\end{figure}

Another 1-loop contribution to $n-\bar n$ oscillations from $\Lambda-\bar \Lambda$ oscillations is given by the diagram of Figure~\ref{fig:app2diagram}. In this diagram, the neutral kaon flavor eigenstates $K^0$ and $\bar K^0$ propagate inside the loop, while the strangeness-changing transition originates from the $K^0-\bar K^0$ mixing term.

The strong-vertex from the term $\Lambda,\, n,\,  K^0$ from the chiral Lagrangian of Eq.~\eqref{ChL} is given by 
\begin{align}
\mathcal{L}_{\text{int}} &= i \,\frac{3F+ D}{f \sqrt 6} \, \bar \Lambda \gamma^\mu \gamma_5 \, \partial_\mu \bar K^0 \, n \, , 
\label{Lambda-strong}
\end{align}
while the second vertex arises from the $K^0 - \bar K^0$ transition, governed by the effective $\Delta S = 2$ Hamiltonian, with $S$ denoting the strangeness quantum number.
At leading order in the electroweak theory, the contribution to $K^0 - \bar K^0$ oscillations arises from box diagrams. The effective Hamiltonian for the $K^0 - \bar K^0$  oscillations can be written as~\cite{Buras:2020xsm}
\begin{equation}
\mathcal{H}_{\text{eff}}^{\Delta S = 2} = \frac{G_F^2 M_W^2}{4\pi^2} \sum_{i,j = c,t} \lambda_i \lambda_j \, \eta_{ij} \, S_0(x_i,x_j) Q^{\Delta S=2} \, ,
\label{Hef}
\end{equation}
with $Q^{\Delta S=2} = [\bar{s} \gamma^\mu 1/2 (1 - \gamma_5) d][\bar{s} \gamma_\mu 1/2(1 - \gamma_5) d]$. Introducing the notation $\braket{\bar K^0|Q^{\Delta S=2} | K^0} = 4/3\, F_K^2\,m_K^2\, \hat B_K$, the transition matrix element between $K^0$ and $\bar K^0$ is then $\Delta m_K=\braket{\bar K^0| \mathcal{H}_{\text{eff}}| K^0} /(2 m_K) $
\begin{eqnarray}
&&\Delta m_K = \frac{G_F^2 M_W^2 F_K^2 m_K}{6 \pi^2} \hat{B}_K  \text{Re} \left[   \lambda_c^2 \eta_1 S_0(x_c)
+\lambda_t^2 \eta_2 S_0(x_t)  + 2 \lambda_c \lambda_t \eta_3 S_0(x_c, x_t)\right] \, ,
\label{DMK1} 
\end{eqnarray}
where $M_W$ is the W boson mass, $F_K=155.7$ MeV is the kaon decay constant~\cite{ParticleDataGroup:2024cfk}, $\lambda_i = V_{is}^* V_{id}$ are CKM factors, $\eta_{i}$ are QCD corrections, $\hat{B}_K$ is the bag parameter, and $S_0(x)$ and $S_0(x_c,x_t)$ are Inami–Lim loop functions. The four-quark operator $Q^{\Delta S=2}(\mu)$ is renormalised at scale $\mu$ in some regularisation scheme, for instance, NDR-$\overline{MS}$. Assuming that
$B_{K}(\mu)$ and the anomalous dimension $\gamma(g)$ are both known in
that scheme, the renormalisation group independent $B$-parameter
$\hat{B}_{K}$ is related to $B_{K}(\mu)$ by the exact formula as given in \cite{FlavourLatticeAveragingGroupFLAG:2024oxs}. 
As in Ref.~\cite{Brod:2011ty} we determine the short distance contribution for the $\Delta m_K \simeq 3.0159 \times 10^{-15}$ GeV, using the latest value of $B_K = 0.717$ \cite{FlavourLatticeAveragingGroupFLAG:2024oxs}. With all ingredients specified, the amplitude for the $n-\bar n$ oscillation induced by the kaon loop is then given by
\begin{eqnarray}
&& \mathcal{M}(n - \bar n)^K  =
 \left(\frac{3F+ D}{f \sqrt 6} \right)^2 \int \, \frac{ d^4 q}{(2 \pi)^4}{v}_n^T C  \slashed{q}  \gamma_5 \frac{1}{ \slashed{q}-m_\Lambda}
   \delta m_\Lambda \frac{1}{ (\slashed{q}-m_\Lambda)} \frac{\Delta m_K2m_K}{(q^2 -m_K^2)^2}  \slashed{q}  \gamma_5 u_n \; ,
 \label{Am-K1}
\end{eqnarray}
 which can be written as
\begin{eqnarray}
&& \mathcal{M}(n - \bar n)^K =- \left(\frac{3F+ D}{f \sqrt 6} \right)^2  \Delta m_K2m_K\, \delta m_\Lambda\,v^T_n C \,\int \, \frac{ d^4 q}{(2 \pi)^4}\frac{ q^2( q^2 + m_\Lambda^2)}{(q^2- m_K^2)^2 (q^2-m_\Lambda^2)^2}\, u_n \; .
\end{eqnarray}
Therefore one finds
\begin{equation}
    \delta m_n^K = \left(\frac{3F+ D
}{f \sqrt 6} \right)^2  \delta m_\Lambda \Delta m_K  2 m_K \int \, \frac{ d^4 q}{(2 \pi)^4}\frac{ q^2( q^2 + m_\Lambda^2)}{(q^2- m_K^2)^2 (q^2-m_\Lambda^2)^2} \, .
\end{equation}
The previous integral is logarithmically divergent and can be computed in Euclidean spacetime using a hard cut-off $\Lambda$, which can be taken to be $\mathcal{O}(1)$ GeV. The result of this integral under this prescription using $\Lambda = 5$ GeV gives
\begin{equation}
    \delta m_n^K \sim 9.8 \times 10^{-16} \; \text{GeV} \, ,
\end{equation}
which translates into the approximate upper bound
\begin{equation}
    \delta m_\Lambda^K\lesssim 1.4 \times 10^{-18} \; \text{GeV} .
\end{equation}

\subsection{Constraints on \texorpdfstring{$\delta m_\Lambda$}{deltamlambda} from dinucleon decay} \label{sec:dinucleon}

\begin{figure}[!htb]
\centering
\includegraphics[scale=1.2]{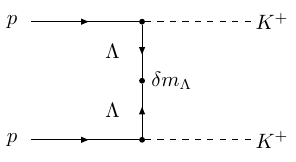}
\caption{\label{fig:ppKK}
Representative tree-level Feynman diagram contributing to the extraction of $\delta m_\Lambda$ from the process $p\,p \to K^+ K^+$ in the $t$-channel. The corresponding $u$-channel diagram is obtained by exchanging the final kaons.
}
\end{figure}

Dinucleon decay is an intranuclear transition in which two nucleons decay into a pair of mesons or leptonic final states. While the latter case has been studied extensively~\cite{Helset:2021plg,He:2021mrt,Gardner:2018azu,Girmohanta:2019cjm,Ma:2025mjy}, this work focuses on decays into two mesons. This channel directly constrains the Wilson coefficients of the dimension-9 SMEFT operators introduced in Sec.~\ref{sec:eftframework}, which are responsible for $|\Delta B|=2$ processes. Experimental searches at Super-Kamiokande have investigated several such processes, including $p\,p \to \pi^+\pi^+$, $n\,n \to 2\pi^0$, and $p\,p \to K^+K^+$, among others~\cite{Super-Kamiokande:2015jbb,Super-Kamiokande:2014hie}.\footnote{For a comprehensive list of BNV processes probed in current and previous experiments, see Ref.~\cite{Heeck:2019kgr}.} Ref.~\cite{Aitken:2017wie} noted that the dinucleon decay $p\,p \to K^+K^+$ can impose particularly stringent bounds on the WC of the $(uds)^2$ operator, the same coefficient that hadronises into $\delta m_\Lambda$. In this section, we employ $B\chi PT$ to determine how the experimental limit from Super-Kamiokande constrains $\delta m_\Lambda$. The amplitude for the $ p\, p \to K^+ \, K^+$ can be calculated from the tree-level diagram presented in Figure~\ref{fig:ppKK}.

\subsubsection*{Calculation of \texorpdfstring{$p \, p \to K^+ K^+$}{ppKK} amplitude}

We can obtain an estimate for the dinucleon decay rate using $B\chi PT$, which provides an estimate of the order of magnitude for such a process induced by the $\delta m_\Lambda$ insertion. In Fig.~\ref{fig:ppKK}, we denote the external proton momenta by $p_1$ (upper line) and $p_2$ (lower line), while the outgoing kaon momenta are $k_1$ (upper line) and $k_2$ (lower line). 

The chiral Lagrangian describing the $p\Lambda K^-$ interaction reads
\begin{align}
\mathcal{L}_{\text{int}} &= i\,\frac{3F + D}{f \sqrt{6}} \, \bar \Lambda\,\gamma^\mu \gamma_5 \, \partial_\mu K^- \, p \;+\; \text{h.c.} \,,
\label{Lambda-pK}
\end{align}
and, in order to evaluate the diagram shown in Fig.~\ref{fig:ppKK}, we also require the corresponding interaction for the charge-conjugated fields:
\begin{equation}
\mathcal{L}_{\text{int}} \;= i\,\frac{3F + D}{f \sqrt{6}} \, \bar{p}^c \gamma^\mu \gamma_5 \, \partial_\mu K^- \,\Lambda^c \;+\; \text{h.c.}\,.
\end{equation}
Here we use the standard definitions $\Psi^c = C \bar{\Psi}^T$ and $\bar{\Psi}^c = -\Psi^T C^\dagger$. The propagator for the charge-conjugated field follows from the relation
\begin{equation}
\left[\sum u^c(p)\, \bar u^c(p) \right]^{-1} \;=\; C \,\frac{\slashed{p} + m}{p^2 - m^2}\, C^\dagger \;=\; \frac{-\,\slashed{p} + m}{p^2 - m^2}\,.
\end{equation}
We use the Mandelstam variables, defined as $s = (p_1 + p_2)^2 = (k_1 + k_2)^2$, $t = (p_1 - k_1)^2 = (k_2 - p_2)^2$, and $u =(p_1 - k_2)^2 = (k_1 - p_2)^2$, with the constraint $s + t + u = 2 m_p^2 + 2 m_K^2$ to write the resulting amplitude for the process $p\,p \to K^+ K^+$ as
\begin{equation}
\mathcal{M}(p\,p \to K^+ K^+) \;=\; 
\left( \frac{3F + D}{f \sqrt{6}} \right)^{2} 
\,\delta m_\Lambda \,
\bar u^{c}(p_{2}) \,\slashed{k}_{1}\slashed{k}_{2} \,
\frac{1}{t - m_\Lambda^{2}} \,
u(p_{1})
\;+\;\{\,t \leftrightarrow u \,,\, k_{1} \leftrightarrow k_{2}\,\} \, .
\label{ppKK-L}
\end{equation}

Note that the previous expression considers free protons, i.e., not bound in nuclear atoms. From this amplitude, one can compute the total rate $\Gamma$ associated with the intranuclear transition, which can be constrained experimentally from the partial lifetime of the decay $^{16}{\rm O} \to {}^{14}{\rm C}\,K^+K^+$. Following the approximate expression given in Ref.~\cite{Goity:1994dq} for the decay into a meson pair,
\begin{equation}
    \Gamma(p\,p \to X) \;\simeq\; \frac{9}{32\pi}\,\frac{\rho_N}{m_N^{2}}\,|\mathcal{M}|^{2} \, ,
    \label{G-nelson}
\end{equation}
where $\rho_N$ denotes the nucleon density. In our analysis, we adopt the average nucleon density $\rho_N \simeq 0.25~{\rm fm}^{-3} \simeq 2 \times 10^{-3}~{\rm GeV}^3$.

To estimate the contribution of Eq.~\eqref{ppKK-L}, we assume that the protons in the oxygen nucleus are nearly at rest, such that $s \simeq 4m_p^2$. For the outgoing kaons, we take $|\vec{k}_1|\simeq|\vec{k}_2| \equiv |\vec{k}| \sim \Lambda_{\rm QCD} \simeq 0.2~{\rm GeV}$, corresponding to a kaon energy $E_K = \sqrt{m_K^2 + |\vec{k}|^2} \simeq 0.53~{\rm GeV}$. With this choice, the Mandelstam variable $t = (p_1-k_1)^2 = (m_p-E_K)^2-|\vec{k}|^2$ evaluates to $t \simeq 0.121~{\rm GeV}^2$. Under these assumptions, the rate becomes
\begin{equation}
    \Gamma(p\,p \to K^+ K^+) \;\simeq\; \frac{9}{32\pi}\,\frac{\rho_N}{m_N^{2}}\,
    \delta m_\Lambda^2\,
    \left( \frac{D + 3F}{f \sqrt{6}} \right)^{4}
    \,\left|\, 2\,\frac{2 m_p }{t - m_\Lambda^2}\,E_K^2 \,\right|^2 \, ,
    \label{G-apr}
\end{equation}
where the factor of 2 multiplying $2m_p$ accounts for the $u$-channel contribution. Using the experimental bound $\tau(p\,p \to K^+K^+) > 1.7 \times 10^{32}$~years~\cite{Super-Kamiokande:2014hie}, corresponding to $\Gamma(p\,p \to K^+K^+) < 1.2 \times 10^{-64}~{\rm GeV}$, we obtain
\begin{equation}
\delta m_\Lambda \;\lesssim\; 1.8 \times 10^{-32}~{\rm GeV} \, .
\label{eq:bounddinucleon}
\end{equation}

Alternatively, one can estimate the rate of the three-body decay $^{16}\mathrm{O} \to {}^{14}\mathrm{C}\,K^+K^+$ directly at the nuclear level. To do so, one must connect the short-distance six-quark interactions discussed in Sec.~\ref{sec:eftframework} with the decay of oxygen nuclei searched for at Super-Kamiokande. Establishing such a connection is extremely challenging, as many of the relevant nuclear inputs 
are not available in the literature. 
Therefore, the following should be regarded as a well-motivated order-of-magnitude estimate.

Since this process occurs inside oxygen nuclei, the corresponding three-body decay rate can be written as
\begin{equation}
    \Gamma(^{16}\mathrm{O} \to {}^{14}\mathrm{C}\,K^+K^+) 
    = \frac{1}{2M_A}\int d\Phi\,|\mathcal{M}_A|^2 \, , 
    \label{eq:dinucleonexpinatom}
\end{equation}
where $M_A$ denotes the mass of the parent nucleus, 
$d\Phi$ the three-body phase-space element, 
and $\mathcal{M}_A$ the corresponding nuclear transition amplitude. 
The latter can be approximated as
\begin{equation}
    \mathcal{M}_A 
    = C_{\mathrm{eff}}\,
      \braket{{}^{14}\mathrm{C}\,K^+K^+|\mathcal{O}_i|{}^{16}\mathrm{O}}
    \;\simeq\;
      \mathcal{C}_i\,
      \braket{K^+K^+|\mathcal{O}_i|pp}\,
      \rho_{pp}(0)
    \;\equiv\;
      \delta m_\Lambda\,\rho_{pp}(0)\,,
\end{equation}
where $\rho_{pp}(0)$ denotes the two-proton contact density within the nucleus. 
Following Ref.~\cite{Weiss:2018zrd}, this quantity can be parameterised as
\begin{equation}
    \rho_{pp}(0) \simeq \kappa\,|\rho_p(0)|^2 \, ,
\end{equation}
where $\rho_p(0) \simeq 0.06 -0.10\;\mathrm{fm}^{-3}$ is the point-proton density in 
$^{16}\mathrm{O}$, and $1 \lesssim \kappa \lesssim 10$ accounts for short-range correlation effects. Assuming, for simplicity, a constant amplitude $\mathcal{M}_A$, 
the corresponding three-body phase-space factor in Eq.~\eqref{eq:dinucleonexpinatom} reads
\begin{equation}
    \int d\Phi
    = \frac{1}{2}\,
      \frac{1}{512\,\pi^3\,M_A^3}
      \int_{s_1}^{s_2}\!ds\,
      \frac{\sqrt{\lambda(M_A^2,M_B^2,s)}\,\sqrt{\lambda(s,m_K^2,m_K^2)}}{s}
    \;\simeq\;
      4\times10^{-7}\,\mathrm{GeV}^{-1}\, ,
\end{equation}
where $M_A$ ($M_B$) denotes the mass of $^{16}\mathrm{O}$ ($^{14}\mathrm{C}$), $s_1 = (2m_K)^2$, $s_2 = (M_A - M_B)^2$ and $\lambda$ is the Källén function. Using the experimental bound $\Gamma^{\mathrm{exp}}(^{16}\mathrm{O} \to {}^{14}\mathrm{C}\,K^+K^+)$, 
together with the central values $\kappa \simeq 4$ and $\rho_p(0) \simeq 0.08\,\mathrm{fm}^{-3}$, one obtains the approximate constraint
\begin{equation}
    \delta m_\Lambda \lesssim 10^{-32}\,\mathrm{GeV}\, ,
\end{equation}
which is of the same order of magnitude as the estimate obtained using the chiral Lagrangian.

It is instructive to compare this result with the bound derived in Ref.~\cite{Aitken:2017wie}, where the authors employed Eq.~\eqref{G-nelson} with $\mathcal{M} \sim \mathcal{C}_{i}\,\langle 2\,{\rm mesons}|\mathcal{O}_{i}|NN\rangle$, 
and estimated $\langle 2\,{\rm mesons}|\mathcal{O}_{i}|NN\rangle \simeq \Lambda_{\rm QCD}^5 \simeq 3.2 \times 10^{-4}~{\rm GeV}^5$. In Ref.~\cite{Aitken:2017wie}, the vacuum insertion approximation was used to estimate $\delta m_\Lambda = \kappa^2\,\mathcal{C}_{i}$,\footnote{In their notation, $\delta_{(uds)^2} \equiv \delta m_\Lambda$. The same approximation was adopted in Ref.~\cite{Arnold:2012sd} to estimate the nuclear matrix element for $n - \bar n$ oscillation.} where one may take $\kappa \simeq 10^{-2}~{\rm GeV}^3$. Using this value yields a slightly weaker bound, $ \delta m_\Lambda \lesssim 2.5 \times 10^{-31}~{\rm GeV}$.

In summary, dinucleon decay yields the strongest constraint on $\delta m_\Lambda$, stronger than both the BESIII $\Lambda-\bar\Lambda$ search and the indirect $n-\bar n$ limit derived above. Its magnitude is comparable to the bounds on $\delta m_n$ shown in Tab.~\ref{tab:experimentalbounds}.

\section{Example: A simplified model} \label{sec:uvmodel}

After introducing the exotic $|\Delta B|=2$ processes in Sec.~\ref{sec:exoticpheno}, we now present a simple UV model featuring a $|\Delta B|=2$ signature. This model serves as a proof of concept, motivating more focused experimental searches for such rare phenomena at current experiments like Super-Kamiokande and future facilities such as Hyper-Kamiokande or DUNE.

We focus on the third model in Tab.~\ref{tab:DeltaB2models}, which was already mentioned in Ref.~\cite{Arnold:2012sd} and more recently in Ref.~\cite{Dorsner:2025pwe} in the context of SU(5) GUT. This model extends the SM particle content by two coloured scalar fields, $\bar S_1 \sim (\mathbf{ \bar 3},\mathbf{1}, -2/3)$ and $\Omega_4 \sim 
(\mathbf{ \bar 6},\mathbf{1}, -4/3)$, with renormalisable couplings specified in Tab.~\ref{tab:definitionsLQ}. Importantly, it avoids the stringent bounds from proton decay searches.\footnote{If heavy right-handed neutrinos $\nu_R$ are present, constraints from the $\Delta(B - L) = 2$ process $p \to K^+ \nu$ push $\Lambda_{\rm BNV}$ to higher scales, raising $M_{\bar S_1}$ to $\mathcal{O}(10^{11})$ GeV~\cite{IBeneito:2025nby}.} Therefore, the set of interactions that lead to $|\Delta B|=2$ processes at tree-level is given by
\begin{equation}
    \mathcal{L} \supset (y^{R}_{\bar{S}_1} )_{[ij]}\, \epsilon_{abc} \,\bar d_{R\, i}^{c \, a} (\bar S_1^*)^b d_{R\,j}^c + (y^{R}_{\Omega_4} )_{\{ij\}}\, \bar u_{R\, i}^{c \, a} (\Omega_4)_{ab} u_{R\,j}^b + \mu  \, (\Omega_4^*)^{ab} \bar S_{1 \,a} \bar S_{1 \,b} +\mathrm{h.c.} \; ,
\end{equation}
where we write explicitly colour indices denoted by $a,b,c$, and $i,j$ denoting quark flavour indices, and we use the same normalisation for the colour sextets as in Ref.~\cite{Arnold:2012sd}.

Constraints from direct searches at colliders set lower bounds on the masses of these scalars around the TeV scale~\cite{CMS:2019gwf}, so we can decouple them at their mass thresholds and be left with the corresponding dimension-9 six-quark operators defined in Sec.~\ref{sec:eftframework}. This is achieved upon integration out at tree-level such heavy degrees of freedom, leading us to the relevant $|\Delta B|=2$ effective operator

\begin{equation} \label{eq:effopmodel}
    \mathcal{L}_{\rm eff}^{|\Delta B|=2} \supset \frac{\mu}{M_{\bar S_1}^4M_{\Omega_4}^2} (y^{R}_{\bar{S}_1} )_{[pq]} (y^{R}_{\bar{S}_1} )_{[rs]} (y^{R}_{\Omega_4} )_{\{tu\}} \epsilon_{abc} \epsilon_{def} (\bar d_{R \, p}^{c \, a} d_{R \, q}^{b})  (\bar d_{R \, r}^{c \, d} d_{R \, s}^{e}) (\bar u_{R \,t}^{c \,c} u_{R \, u}^{f}) +\mathrm{h.c.} \, .
\end{equation}

Note that due to the antisymmetric nature of the coupling $y^{R}_{\bar{S}_1}$, the previous operator does not induce $n - \bar n$ oscillations at tree level.\footnote{The role of $\bar S_1$ in forbidding tree-level $n - \bar n$ oscillations was noted early in the literature~\cite{Basecq:1983hi}.} In this section, we examine the $|\Delta B| = 2$ phenomena arising in this scenario, focusing in particular on $\Lambda - \bar \Lambda$ oscillations and dinucleon decay such as $p \,p \to K^+ K^+$.

\subsection{\texorpdfstring{$|\Delta B|=2$}{deltab2} phenomenology} \label{sec:modelpheno}

This model exemplifies the generation of $\delta m_\Lambda$ at tree-level from the effective Lagrangian of Eq.~\eqref{eq:effopmodel}. In the following, we focus on each of the different $|\Delta B|=2$ processes separately:

\subsubsection*{\texorpdfstring{$n - \bar n$ and $\Lambda - \bar \Lambda$}{ll} oscillations}

\begin{figure}[htp]
    \centering
    \includegraphics[width = 8cm]{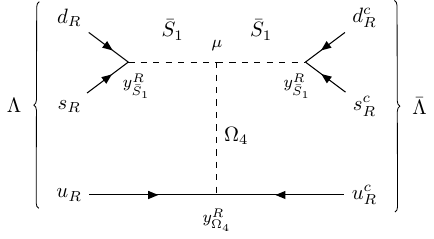}
    \caption{Feynman diagram leading to $\Lambda - \bar \Lambda$ at tree-level in the model described in Sec.~\ref{sec:uvmodel}. See main text for details.}
    \label{fig:LLOandNNOmodel}
\end{figure}

We illustrate the tree-level contribution to $\Lambda - \bar \Lambda$ oscillations in this model in Figure~\ref{fig:LLOandNNOmodel}, where the antisymmetric flavour structure of the coupling $y^{R}_{\bar{S}_1}$ naturally enables this process.

\begin{figure}[ht]
    \centering
    \includegraphics[width = 11cm]{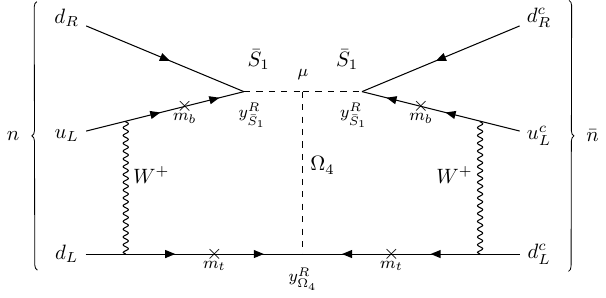}
    \caption{Feynman diagram leading to $n-\bar n$ oscillations at 2 loops in the model described in Sec.~\ref{sec:uvmodel}. The diagram features third-family quarks $b$ and $t$, although other quark flavors (except $d$) may also contribute. See main text for details.}
    \label{fig:2loops mediated NNO}
\end{figure}

Our objective is to demonstrate that $\Lambda - \bar \Lambda$ oscillations dominate over $n - \bar n$ oscillations in current and future experimental searches, despite that $\delta m_n^{\rm exp}/\delta m_
\Lambda^{\rm exp}\simeq 6 \times 10^{-16}$, as summarized in Tab.~\ref{tab:experimentalbounds}. This apparent tension is resolved by noting that, within this model, $\Lambda - \bar \Lambda$ oscillations are generated at tree level, whereas $n - \bar n$ oscillations arise only at two-loop order, via the diagram shown in Figure~\ref{fig:2loops mediated NNO}, leading to a highly suppressed mass mixing $\delta m_n$.\footnote{Ref.~\cite{Dorsner:2025pwe} also discusses an analogous two-loop topology responsible for $n - \bar n$ oscillations.}

We can use the diagram of Fig.~\ref{fig:2loops mediated NNO} to estimate the loop-level contribution to the WC responsible for $n - \bar n$ oscillations $\mathcal{C}_i^{n - \bar n}$. This contribution is strongly suppressed by CKM-angle factors, mass insertions, and the two-loop factor. In this specific model, $\mathcal{C}_i^{n - \bar n}$ can be estimated as
\begin{align} \label{eq:generalestimateLLO}
    \mathcal{C}_i^{\, n - \bar n} \sim & \left(\frac{\mu}{M_{\bar S_1}^4 M_{\Omega_4}^2} \right)\, G_F^2 \frac{1}{(16\pi^2)^2} \times \nonumber \\ & \sum_{p,\,q=2}^{3}\sum_{r,\,s=1}^{3} \left( y_{\bar S_1}^R\right)_{1p} \left( y_{\bar S_1}^R\right)_{1q} \left(y_{\Omega_4}^R\right)_{rs}[m_{d}]^p[m_d]^q[m_u]^r[m_u]^s V^*_{1q}V^*_{1p} V_{r1}V_{s1} \; ,
\end{align}
where the indices $p,q$ $(r,s)$ label the down-type (up-type) quark flavours propagating in the loop. The symbols $[m_d]^p$ and $[m_u]^p$ are the masses of the $p$-th family down-type and up-type quarks, respectively, and $V_{pq}$ is the corresponding CKM matrix element. We have neglected the logarithms of the running from the TeV scale down to neutron mass. The antisymmetry of $y_{\bar S_1}^R$ forbids the choice $r,s = 1$. Interestingly, Eq.~\eqref{eq:generalestimateLLO} exhibits a GIM-like cancellation mechanism, whereby large quark masses are compensated by small mixing angles. As a result, the contributions from strange and bottom quarks circulating in the loop are of comparable magnitude. For definiteness, we consider the case in which top and bottom quarks run inside the loop as depicted in Fig.~\ref{fig:2loops mediated NNO}, obtaining a rough estimate of
\begin{equation}
    \mathcal{C}_i^{\,n - \bar n} \sim \left(\frac{G_F^2}{(16\pi^2)^2} m_t^2m_b^2(V_{ub}^*)^2V_{td}^2 \right) \, \frac{\mu}{M_{\bar S_1}^4 M_{\Omega_4}^2} \; \sim 10^{-18} \, \frac{1}{\Lambda_{\rm BNV}^5} \; ,
\end{equation}
where we have assumed the Yukawa couplings $y_{\bar S_1}^R$, and $y_{\Omega_4}^R$ of Eq.~\eqref{eq:generalestimateLLO} to be $\mathcal{O}(1)$, and we have defined a new effective BNV scale $\Lambda_{\rm BNV}$ as
\begin{equation} \label{eq:effbnvscale}
    \Lambda_{\rm BNV}^5 = \frac{M_{\bar S_1}^4M_{\Omega_4}^2}{\mu} \, .
\end{equation}

Using the experimental lower bounds on $\tau_{\Lambda - \bar \Lambda}$ and $\tau_{n - \bar n}$ of Tab.~\ref{tab:experimentalbounds}, we find two lower bounds on $\Lambda_{\rm BNV}$ coming from each observable:
\begin{enumerate}
    \item $\mathbf{\Lambda - \bar \Lambda}$: The tree-level diagram in Fig.~\ref{fig:LLOandNNOmodel} generates $\Lambda - \bar \Lambda$ oscillations, yielding a WC of the order
    \begin{equation}
        \mathcal{C}_i^{\, \Lambda - \bar \Lambda} \sim \frac{1}{\Lambda_{\rm BNV}^5} \, .
    \end{equation}
    If we introduce the experimental sensitivities from BESIII~\cite{BESIII:2024gcd} and estimate the unknown nuclear matrix element $\braket{\bar \Lambda|\mathcal{O}|\Lambda}\sim10^{-5}$ GeV$^6$, we find from Eq.~\eqref{eq:defdeltam}
    \begin{equation} \label{eq:ll0modelestimate}
    \braket{\bar \Lambda|\mathcal{O}|\Lambda} \; \frac{1}{\Lambda_{\rm BNV}^5} \lesssim 10^{-18} \; \rm GeV \implies \frac{1}{\Lambda_{\rm BNV}^5} \lesssim 10^{-13} \; \rm GeV \; .
    \end{equation}
    \item {\boldmath$ n - \bar n$}: In this model we generate $n - \bar n$ oscillations at two-loops through the diagram in Fig.~\ref{fig:2loops mediated NNO}, whose contribution to $n - \bar n$ oscillations can be estimated as
    \begin{equation}
        \mathcal{C}_i^{\, n - \bar n} \sim 10^{-18}\frac{1}{\Lambda_{\rm BNV}^5} \, .
    \end{equation}
    If we introduce the experimental sensitivities from Super-K and use the nuclear matrix element $\braket{\bar n|\mathcal{O}|n}\sim10^{-5}$ GeV$^6$ computed in the lattice~\cite{Rinaldi:2018osy,Rinaldi:2019thf}, we find from Eq.~\eqref{eq:defdeltam}
    \begin{equation} \label{eq:nno0modelestimate}
    10^{-18} \braket{\bar n|\mathcal{O}|n}\frac{1}{\Lambda_{\rm BNV}^5} \lesssim 10^{-33} \; \rm GeV \implies \frac{1}{\Lambda_{\rm BNV}^5} \lesssim 10^{-10} \; \rm GeV \; .
    \end{equation}
\end{enumerate}
Therefore, from Eqs.~\eqref{eq:ll0modelestimate} and \eqref{eq:nno0modelestimate}, we see that current sensitivities from $\Lambda - \bar \Lambda$ put a stronger bound on the parameters of the UV theory. From the former expression of Eq.~\eqref{eq:ll0modelestimate}, we find that \begin{equation} \label{eq:currentbounLL0}
    \Lambda_{\rm BNV} \gtrsim 400 \; \rm GeV \, ,
\end{equation}
or explicitly in terms of all parameters of the UV theory
\begin{equation}\label{eq:benchmarkvaluesLLOModel}
    \left(\frac{M_{\bar S_1}}{400 \; \rm GeV}\right)^4\left(\frac{M_{\Omega_4}}{400 \; \rm GeV}\right)^2\left(\frac{400 \; \rm GeV}{\mu}\right) \gtrsim \frac{\left(y_{\bar S_1}^R\right)^2 y_{\Omega_4}^R}{1} \left(\frac{\braket{\bar \Lambda|\mathcal{O}|\Lambda}}{10^{-5} \; \rm GeV^6}\right) \; .
\end{equation}
The lower mass bounds for $\bar{S}_1$ and $\Omega_4$ derived from $\Lambda-\bar \Lambda$ oscillations are about an order of magnitude weaker than the few-TeV constraints from direct collider searches~\cite{CMS:2019gwf}, making the current BESIII limits non-competitive.

For the BSM process of $\Lambda - \bar \Lambda$ oscillations to provide stronger or competitive constraints on the WCs compared to the well-constrained $n- \bar n$ oscillations, the $n - \bar{n}$ amplitude must be suppressed. This can be achieved through mechanisms such as the two-loop generation of $n - \bar{n}$ oscillations, or from the exchange of heavy $W$ bosons in internal loops, which effectively weaken the impact of the stringent experimental bounds presented in Tab.~\ref{tab:experimentalbounds}.\footnote{Dominant $\Lambda - \bar \Lambda$ oscillations relative to $n - \bar{n}$ can also arise by enforcing a flavour symmetry in the UV theory, such as an approximate $\mathrm{U}(3)^5$~\cite{DAmbrosio:2002vsn} or $\mathrm{U}(2)^5$ flavour symmetry~\cite{Barbieri:2011ci,Barbieri:2012uh}. Under such flavour structures, couplings involving first-generation fermions are typically more suppressed than those involving second- and third-generation fermions~\cite{Baldes:2011mh,Dong:2011rh}.}

It is worth commenting on the feasibility of achieving such suppression mechanisms within tree-level completions of the $|\Delta B| = 2$ SMEFT operators. Among the scalar-mediated tree-level models for $|\Delta B| = 2$ transitions shown in Tab.~\ref{tab:DeltaB2models}, only one scenario yields stronger bounds from $\Lambda - \bar \Lambda$ oscillations. The alternative tree-level topology involving both a scalar and a heavy fermion, which is described in Appendix~\ref{sec:fermionscalartopology}, does not substantially increase the number of viable models. For instance, the scalar $\bar S_1$ and the fermion with quantum numbers $(\mathbf{8},\mathbf{1},0)_F$ would lead to $\Lambda - \bar \Lambda$ oscillations more dominantly through the same suppression than the model explained above. Additional possibilities emerge at the loop level, where $|\Delta B| = 2$ transitions may arise through a richer variety of mechanisms.

\subsubsection*{\texorpdfstring{$p \, p \to K^+ K^+$}{dnd} dinucleon decay}

\begin{figure}[ht]
\centering
\includegraphics[scale=1]{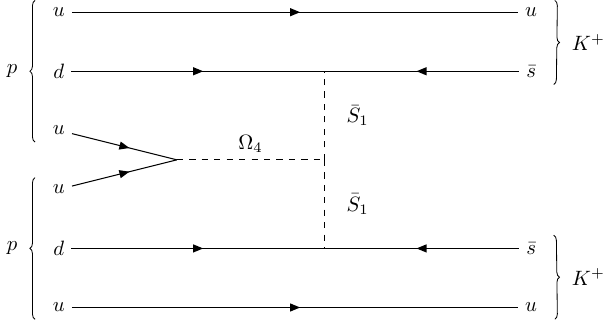}
\caption{\label{fig:dinucleonUVtheory}Dinucleon decay $p\,p\to K^+K^+$ arising at tree-level in the model described in Sec.~\ref{sec:uvmodel}. See main text for details.}
\end{figure}

The model presented above also generates $|\Delta B|=2$ dinucleon decays such as $p\,p \to K^+K^+$ at tree level. Such dinucleon decays are actively searched for in Oxygen nuclei within water molecules in experiments like Super-Kamiokande, which establishes the most stringent bound for the process $^{16}O \to ^{14}C \;K^+K^+$ of $\tau > 1.7 \times 10^{32}$ years~\cite{Super-Kamiokande:2014hie}. Such a BSM phenomenon can be generated through the tree-level diagram depicted in Figure~\ref{fig:dinucleonUVtheory}.

We use the tree-level diagram in Fig.~\ref{fig:dinucleonUVtheory} for $ p \,p \to K^+ K^+ $ to derive a bound on the BNV scale $\Lambda_{\rm BNV}$ defined in Eq.~\eqref{eq:effbnvscale}, in direct analogy to the $n - \bar n$ oscillation estimate of Eq.~\eqref{eq:roughestimatennbar}. In our model, the estimate for $\delta m_\Lambda$ is
\begin{equation}
    \delta m_\Lambda \sim \mathcal{C}_i^{\,\Lambda-\bar \Lambda}\braket{\bar \Lambda|\mathcal{O}|\Lambda} \sim \frac{1}{\Lambda_{\rm BNV}^5}\braket{\bar \Lambda|\mathcal{O}|\Lambda}\; .
\end{equation}
Using the bound from dinucleon decay
$\delta m_\Lambda \lesssim 10^{-32}$ GeV derived in Sec.~\ref{sec:dinucleon} using $B \chi PT$, and estimating $\braket{\bar \Lambda|\mathcal{O}|\Lambda} \sim 10^{-5}$ GeV$\,^6$, one obtains
\begin{equation} \label{eq:benchmark dinucleonmodel}
    \Lambda_{\rm BNV} \gtrsim 300 \; \rm{TeV} \, .
\end{equation}

To conclude, we write in Table~\ref{tab:modelbounds} the summary of the different $|\Delta B|=2$ processes observable in this model and the lower bound obtained for $\Lambda_{\rm BNV}$ from each process. The data clearly show that current searches for dinucleon decay, particularly $p\, p \to K^+ K^+$, place the most powerful constraints on BNV in this model.

\begin{table}[ht]

\centering
\renewcommand{\arraystretch}{1.5}
\setlength{\tabcolsep}{14pt}

\begin{tabular}{|c|c|c|c|}
\hline
\textbf{ \boldmath$|\Delta B|=2$ process} & \textbf{Loop order} & \textbf{Experiment} & \textbf{Lower bound \boldmath$\Lambda_{\rm BNV}$ (TeV)} \\ 
\hline
\hline

$n - \bar n$ & 2-loops & Super-K~\cite{Super-Kamiokande:2020bov} & $0.1$ \\[0.1em] \hline 

$\Lambda - \bar \Lambda$ & Tree-level & BESIII~\cite{BESIII:2024gcd} & $ 0.4$ \\[0.1em] \hline

$p \, p \to K^+K^+$ & Tree-level& Super-K~\cite{Super-Kamiokande:2014hie} & $300$ \\[0.1em] \hline
\end{tabular}
\caption{Lower bounds on the scale $\Lambda_{\rm{BNV}}$ from the different $|\Delta B|=2$ processes generated by the model of Sec.~\ref{sec:uvmodel}. The first column lists the $|\Delta B|=2$ process, the second gives the leading loop order, the third specifies the relevant experimental search, and the fourth states the derived lower limit on $\Lambda_{\rm BNV}$.}
\label{tab:modelbounds} 
\end{table}

For completeness, it is also noteworthy to consider constraints from collider searches and flavour physics, which we will briefly address.

\subsubsection*{Constraints from collider searches}

Current LHC searches place stringent lower bounds on the masses of the coloured scalars $\bar S_1$ and $\Omega_4$, excluding values below 1.8 TeV~\cite{CMS:2019gwf}. As we already mentioned above, these limits are well beyond the mass scales that can be probed in present searches for $\Lambda - \bar \Lambda$ oscillations at BESIII~\cite{BESIII:2024gcd} as it can be seen in Tab.~\ref{tab:modelbounds}. It is therefore reasonable to ask what level of sensitivity future experiments would need to achieve in order for indirect bounds from $\Lambda - \bar \Lambda$ oscillations to become competitive with direct collider searches. In fact, it has been suggested in the literature that upcoming $\tau$–charm factories may reach sensitivities of order $\delta m_\Lambda \lesssim \mathcal{O}(10^{-20})$ GeV~\cite{Addazi:2020nlz,Kang:2009xt,Achasov:2023gey}.

For illustration, we consider the benchmark choice 
$M_{\bar S_1} \simeq M_{\Omega_4} \simeq 2$ TeV and 
$\mu \simeq 2$ TeV. In this case, one estimates
\begin{equation}
\delta m_\Lambda = \mathcal{C}_i \braket{\bar \Lambda | \mathcal{O}_i| \Lambda}
\simeq \frac{\mu}{M_{\bar S_1}^4 M_{\Omega_4}^2} \braket{\bar \Lambda | \mathcal{O}_i| \Lambda}
\simeq 10^{-22}~\text{GeV} \,.
\end{equation}
Hence, for oscillation searches such as the ones carried at BESIII to become competitive with collider limits, future experiments would need to improve the present bounds on $\delta m_\Lambda$ by approximately four orders of magnitude, a level of precision that is infeasible in any foreseeable future.

\subsubsection*{Constraints from flavour physics}

Note that flavour-changing neutral currents (FCNCs) are very much suppressed in the SM, in particular $D^0 - \bar D^0$, which puts stringent constrains on the combination~\cite{Chen:2009xjb}
\begin{equation} \label{eq:ddmixing}
    (y_{\Omega_4}^R)_{11}(y_{\Omega_4}^R)_{22}^* \lesssim\left( \frac{M_{\Omega_4}}{10^4 \; \rm TeV}\right)^2\; ,
\end{equation}
which could push further into the UV the mass of the colour sextet, thus rendering non-competitive the phenomenological constraint from $p \, p \to K^+K^+$ of Eq.~\eqref{eq:benchmark dinucleonmodel}. It is worth noting, however, that the bound in Eq.~\eqref{eq:ddmixing} applies only to $M_{\Omega_4}$, since $\bar S_1$ does not induce FCNCs owing to the antisymmetric nature of its coupling to right-handed down-type quarks. Moreover, Eq.~\eqref{eq:ddmixing} constrains a different set of couplings from those relevant to $|\Delta B| = 2$ observables. Under the simplifying assumption that all couplings are $\mathcal{O}(1)$, one can nonetheless use Eq.~\eqref{eq:ddmixing} to relax the lower bound on the mass of $\bar S_1$ from 300 TeV, as given in Eq.~\eqref{eq:benchmark dinucleonmodel}, to $M_{\bar S_1} \gtrsim 30$ TeV. This estimate assumes $M_{\Omega_4} \gtrsim 10^4$ TeV and $\mu \simeq M_{\bar S_1}$, corresponding to the least restrictive scenario for $M_{\bar S_1}$.

\begin{figure}[ht]
\centering
\includegraphics[scale=0.7]{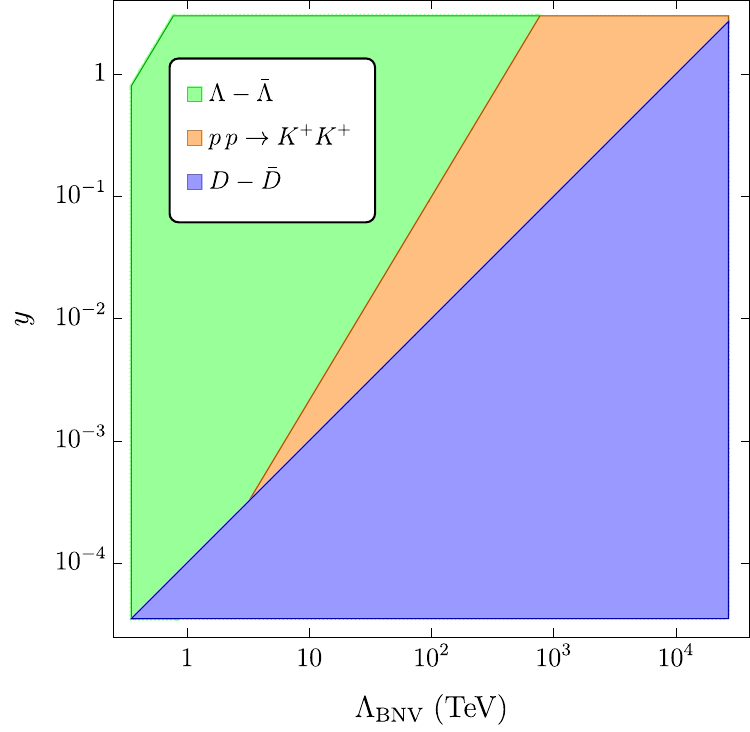}
\caption{\label{fig:parameterspace} Parameter space of the model combining the experimental constraints from $\Lambda-\bar \Lambda$ oscillations (green), dinucleon decay bounds (orange), and $D-$meson oscillations (blue). One can easily see that current constraints from BESIII experiment are not competitive with the rest of the experiments. See main text for more details.}
\end{figure}

In Figure~\ref{fig:parameterspace}, we combine the three kinds of constraints on the effective scale $\Lambda_{\rm BNV}$, and for simplicity, all dimensionless couplings to be equal, i.e. $y \equiv (y_{\Omega_4}^{\bar S_1})_{ij} =(y_{\Omega_4}^R)_{ij}$. We are aware that such considerations are oversimplifying, but for illustrative purposes, it's enough to do it like this. Therefore, one may conclude that current bounds from $\Lambda - \bar \Lambda$ oscillations are far from being competitive with other kinds of bounds.

\section{Conclusions}\label{sec:conclusions}

Processes that violate baryon number constitute one of the most compelling avenues for uncovering physics beyond the SM.  Such processes naturally emerge in GUTs, where various BNV patterns can occur. 
Canonical examples include proton decay, which violates baryon number by one unit, and neutron–antineutron oscillations, which violate it by two units. 
The latter provides one of the most extensively studied experimental probes of $|\Delta B| = 2$ transitions. 
However, since $n-\bar{n}$ oscillations involve only first-generation fermions, they constrain only a limited subset of UV completions capable of generating such operators. 
To explore a broader class of $|\Delta B| = 2$ interactions, it is therefore important to consider more exotic processes, such as $\Lambda-\bar \Lambda$ oscillations and dinucleon decays like $p\,p \to K^+ K^+$, which probe distinct combinations of WCs and may, in some cases, be experimentally more accessible than conventional $n-\bar{n}$ searches. In this work, we focus on these exotic processes and perform a systematic and comprehensive study of their phenomenology.

Building on this motivation, we revisit BNV processes from complementary perspectives, both within an EFT framework and from the standpoint of a UV-complete theory. In each case, we employ the phenomenological Chiral Lagrangian to compute the branching ratios associated with these processes and derive bounds on the relevant New Physics scale $\Lambda_{\rm BNV }$, as well as on the mass splitting $\delta m$ responsible for $|\Delta B| = 2$ particle oscillations.

Our analysis demonstrates that, despite ongoing and future efforts to improve the limits on $n-\bar n$ oscillations, there exist models that are more tightly constrained by other exotic $|\Delta B| = 2$ phenomena, such as $\Lambda-\bar \Lambda $ oscillations and dinucleon decays $p\,p \to K^+ K^+$. Our analysis shows that dinucleon decay searches, particularly those at Super-Kamiokande, already set the most stringent limits on strangeness-violating $|\Delta B| = 2$ processes, probing scales far beyond the reach of current and planned $\Lambda-\bar \Lambda$ oscillation experiments.

\begin{table}[ht]

\centering
\renewcommand{\arraystretch}{1.5}
\setlength{\tabcolsep}{11pt}

\begin{tabular}{|c|c|c|c|}
\hline
\textbf{\boldmath $|\Delta B|=2$ process} & \textbf{Order in \boldmath $B\chi PT$} & \textbf{Experiment} & \textbf{Upper bound \boldmath $\delta m_\Lambda$ (GeV)} \\ 
\hline
\hline

$\Lambda - \bar \Lambda$ & Tree-level & BESIII~\cite{BESIII:2024gcd} & $ 10^{-18}$ \\[0.1em] \hline

\multirow{3}{*}{$n - \bar n$} & Tree-level & \multirow{3}{*}{Super-K~\cite{Super-Kamiokande:2020bov}} & $10^{-20}$ \\[0.1em]
& 1-loop ($K$) & & $10^{-18}$ \\[0.1em]
& 1-loop ($\pi$) & & $10^{-18}$ \\[0.1em]\hline 

$p \, p \to K^+K^+$ & Tree-level& Super-K~\cite{Super-Kamiokande:2014hie} & $10^{-32}$ \\[0.1em] \hline
\end{tabular}
\caption{Summary of current experimental upper bounds on the mass splitting $\delta m_\Lambda$ for various $|\Delta B| = 2$ processes. The first column lists the process under consideration, and the second indicates the order at which it appears in $B\chi$PT. In the third column, we specify the corresponding experimental search providing the bound, and in the fourth, we provide the resulting upper limit on $\delta m_\Lambda$ in GeV.}
\label{tab:conclusions} 
\end{table}

For this reason, we urge the experimental community to pursue more dedicated searches for dinucleon decay, revisiting the existing Super-Kamiokande limits, which have remained unchanged for over a decade, and to exploit upcoming opportunities at detectors such as Hyper-K and DUNE. These channels may offer the most promising path toward discovering baryon number violation in more exotic sectors. Likewise, we encourage lattice QCD collaborations to compute the relevant matrix elements needed to precisely match the LEFT operators onto the hadronic-level calculations of $|\Delta B| = 2$ processes, including both oscillation rates and dinucleon decay amplitudes.

In conclusion, the observation of BNV would unequivocally signal a new era in fundamental physics. The discovery of exotic $|\Delta B| = 2$ processes beyond $n - \bar n$ oscillations would open new avenues for exploration and point unequivocally to physics beyond minimal scenarios. The intricate patterns revealed by such a discovery could find their UV origin in the results presented in this work.

\acknowledgments

We are deeply grateful to the organisers of the Workshop ``INT-25-91W: Baryon Number Violation: From Nuclear
Matrix Elements to BSM Physics''~\cite{Broussard:2025opd}, which brought us together and inspired this work. ABB thanks Juan Herrero García and Arcadi Santamaria for insightful discussions, and to the Jo\v zef Stefan Institute for their warm hospitality during his enjoyable visit, when part of this work was carried out. ABB is funded by the grant CIACIF/2021/061 of the ``Generalitat Valenciana'' and also by the Spanish ``Agencia Estatal de Investigación'' through MICIN/AEI/10.13039/501100011033. Feynman diagrams were
generated using the Ti\textit{k}Z-Feynman package for
\LaTeX~\cite{Ellis:2016jkw}. SF acknowledge the financial support from the Slovenian Research Agency (research core funding No. P1-0035 and N1-0321). AAP was supported in part by the US Department of Energy grant DE-SC0024357.

\appendix

\section{UV completions involving exotic fermions}
\label{sec:fermionscalartopology}

\begin{figure}[ht]
    \centering
    \includegraphics[width = 7.5cm]{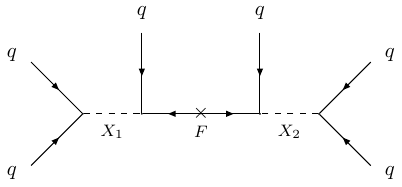}
    \caption{Scalar-fermion topology leading to $|\Delta B|=2$ processes at tree-level. The label $q$ denotes generic quark-type fermions. See main text for further details.}
    \label{fig:fermionUVtopology}
\end{figure}

In this appendix, we explore the generation of $|\Delta B| = 2$ processes at tree level, mediated by one or two exotic scalars, defined in Tab.~\ref{tab:definitionsLQ}, and a massive exotic fermion. These interactions proceed through the topology shown in Figure~\ref{fig:fermionUVtopology}.

In contrast to the purely scalar UV completions discussed in Sec.~\ref{sec:3scalars}, which follow the topology shown in Fig.~\ref{fig:scalarUVtopology} and yield only nine possibilities, we identify 37 distinct models involving two exotic scalars, $X_1$ and $X_2$, and one heavy fermion, $F$. Such UV completions have received little attention in the literature. After completing our analysis, we found that they had previously been considered only in Refs.~\cite{Chen:2022gjd,Heeck:2026dmh}. Our results, obtained independently, are in full agreement with those references.

Specific UV models corresponding to this topology were discussed in some of the early works on the subject~\cite{Zwirner:1983dgv,Barbieri:1985ty,Mohapatra:1986bd}, as well as in more recent works~\cite{Dorsner:2025epy,Dev:2015uca}.\footnote{The earliest study of $n - \bar{n}$ oscillations~\cite{Kuo:1980ew} involved the exchange of a right-handed neutrino $N_R$, corresponding to the topology discussed in this section. However, the mediator in that case was a gauge boson, as GUTs were the prevailing framework at the time.} To obtain the full list of such UV completions, we employed the \textsc{Mathematica} package \texttt{Sym2Int}~\cite{Fonseca:2017lem,Fonseca:2019yya}. The tree-level classification through this topology is provided in Table~\ref{tab:fermionUVtopology}.

\begin{table}[ht]
\centering

\setlength{\arrayrulewidth}{0.2mm}
\setlength{\tabcolsep}{6pt}
\renewcommand{\arraystretch}{1.5}
    
\large
\begin{tabular}{|c|c|c|c|c|c|c|c|}
    \hline
       
    \diagbox[width=3.5em,height=3.5em]{\boldmath $X_1$}{\boldmath $X_2$}& \boldmath $S_1$ \xmark & \boldmath $S_3$ \xmark &  \boldmath $\bar S_1$ & \boldmath $\Omega_1$ & \boldmath $\Upsilon$ & \boldmath $\Omega_2$ & \boldmath $\Omega_4$ \\ \hline  \hline 
    
    \multirow{2}{*}{\boldmath $S_1$ \xmark} & \multirow{2}{*}{$N_R, \;F_N$}  & & $\Delta_1, \; F_{\Delta_1}, \; E $& \multirow{2}{*}{$F_N$} & & \multirow{2}{*}{$ F_{\Delta_1}, \; F_E, \; F_N $} & \\ 
    & & & $F_E, \; N_R, \; F_N  $  & & & & \\
    \hline

    \boldmath $S_3$ \xmark
    & \cellcolor{gray!20} & $\Sigma, \; F_\Sigma $& $\Delta_1, \; F_{\Delta_1}$ & & $F_\Sigma$ & $F_{\Delta_1}$ & \\ \hline

    \boldmath $\tilde S_1$ \xmark
    & \cellcolor{gray!20} & \cellcolor{gray!20} & $E, \; F_E $ & & & $F_E $ & \\ \hline
       
    \boldmath $\bar S_1$
    & \cellcolor{gray!20} & \cellcolor{gray!20} & $N_R $ \xmark, $F_N$ & $ F_{\Delta_1}, \; F_N, \; F_E $ & $F_{\Delta_1}$ & $F_N$ & $F_E $ \\ \hline

    \boldmath $\Omega_1$
    & \cellcolor{gray!20} & \cellcolor{gray!20} & \cellcolor{gray!20} & $F_N$ & & $F_{\Delta_1}, \; F_E, \; F_N$ &  \\ \hline

    \boldmath $\Upsilon$
    & \cellcolor{gray!20} & \cellcolor{gray!20} & \cellcolor{gray!20} & \cellcolor{gray!20} & $F_\Sigma $ & $F_{\Delta_1}$ & \\ \hline

    \boldmath $\Omega_2$
    & \cellcolor{gray!20} & \cellcolor{gray!20} & \cellcolor{gray!20} & \cellcolor{gray!20} & \cellcolor{gray!20} & $F_N $ & $ F_E$ \\ \hline

    \end{tabular}
       
\caption{Tree-level UV completions of $|\Delta B|=2$ SMEFT operators via the fermion topology in Fig.~\ref{fig:fermionUVtopology}, extending the SM by two exotic scalars and a fermion. Each entry lists the heavy fermion $F$ that, together with the bold scalar $X_1$ (rows) and $X_2$ (columns), completes the model. Multiple fermion options exist for some scalar pairs. Symmetric combinations under $X_1 \leftrightarrow X_2$ are shaded grey. Models excluded by stringent proton decay limits are marked with \xmark.}
\label{tab:fermionUVtopology}
\end{table}

In general, these completions require a fermion and either two distinct scalar fields or, in the minimal setup, a single scalar. Remarkably, all minimal completions of this type involve a fermion with zero hypercharge. By examining the quantum numbers of the exotic fermions that appear in these completions, we find that $|\Delta B| = 2$ SMEFT operators admit UV completions involving standard type-I and type-III seesaw particles responsible for the tree-level UV completion of the Weinberg operator~\cite{Weinberg:1979sa}, namely $N_R \sim (\mathbf{1},\mathbf{1},0)$ and $\Sigma \sim (\mathbf{1},\mathbf{3},0)$, respectively, as well as vector-like copies of SM leptons: $E \equiv E_L + E_R \sim (\mathbf{1},\mathbf{1},-1)$ and $\Delta_1 \equiv \Delta_{1L} + \Delta_{1R} \sim (\mathbf{1},\mathbf{2},-1/2)$.
For the type-I and type-III seesaw mediators, similar UV completions have been discussed previously~\cite{Dorsner:2025pwe,Berezhiani:2015afa,Dorsner:2025epy,Grojean:2018fus}. Nevertheless, because these UV completions contain scalar fields inducing proton decay at tree level, the stringent proton lifetime limits make them phenomenologically unviable, with the associated $\delta m$ for $|\Delta B| = 2$ processes being strongly suppressed.\footnote{In the case of the $\bar S_1$ and $N_R$ mediators, the model generates a $|\Delta (B-L)| = 2$ tree-level proton decay channel via a dimension-7 SMEFT operator~\cite{Beneito:2023xbk}, which is also subject to stringent experimental constraints.}

On the other hand, exotic fermions that transform as colour octets under $\rm SU(3)_C$ lie outside the so-called Linear SM Extensions (LSMEs)~\cite{IBeneito:2025nby} and thus do not appear in the general tree-level dictionary of SM extensions onto dimension-6 operators given in Ref.~\cite{deBlas:2017xtg}. We denote these fields as follows: 
\begin{equation*}
    F_N \sim (\mathbf{8},\mathbf{1},0), \quad F_{\Delta_1} \sim (\mathbf{8},\mathbf{2},\frac{1}{2}), \quad F_\Sigma \sim (\mathbf{8},\mathbf{3},0), \quad F_E \sim (\mathbf{8},\mathbf{1},1) \, ,
\end{equation*}
where the subscript refers to the corresponding colour-singlet fermion with the same $\rm SU(2)_L \times U(1)_Y$ quantum numbers. Some of these colour-octet fermions participate, in conjunction with other fields, in tree-level completions of dimension-8 SMEFT operators~\cite{Li:2023pfw}, or appear in loop-level completions of dimension-6 BNV SMEFT operators, contributing to processes such as proton decay~\cite{Helo:2019yqp}.

\section{UV completions with quartic couplings} \label{sec:4scalars}

\begin{figure}[ht]
    \centering
    \includegraphics[width = 5.5cm]{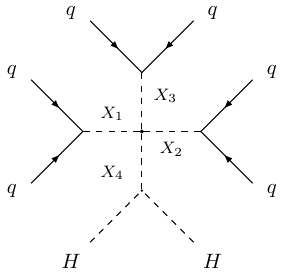}
    \caption{Schematic representation of the quartic interaction $\lambda, X_1 X_2 X_3 X_4$ responsible for $|\Delta B| = 2$ transitions at tree level. Here, $q$ denotes generic quark-type fermions. Two Higgs fields $H$ are shown on the scalar external legs, although $X_4$ can in general couple to three possible combinations: the $\mathrm{SU(2)_L}$ triplet $HH$, or the singlet/triplet structures $H^\dagger H$. See main text for further details.}
    \label{fig:quarticUVtopology}
\end{figure}

In this appendix, we investigate the possibility of generating $|\Delta B| = 2$ transitions through quartic couplings in the UV, through the topology depicted in Figure~\ref{fig:quarticUVtopology}.

In this new possibility, the quartic coupling $\lambda \, X_1X_2X_3X_4$ involves always four exotic scalars, three of them fall in the category of scalars with diquark couplings listed in Tab.~\ref{tab:definitionsLQ}, and the last scalar $X_4$ can only take three possibilities, given by $X_4 = \lbrace \Xi_1 \sim (\mathbf{1},\mathbf{3},1), \; \Xi_0\sim (\mathbf{1},\mathbf{3},0), \; \mathcal{S}\sim (\mathbf{1},\mathbf{1},0)\rbrace$, following the notation of Ref.~\cite{deBlas:2017xtg}.

This mechanism, which also gives rise to $|\Delta B| = 2$ processes, has been largely overlooked in the literature, mainly because the leading contributions to such phenomena are expected to originate from effective operators of lower mass dimension, namely those of dimension~9. However, there exist UV models that do not generate $|\Delta B| = 2$ processes at tree level via dimension-9 operators, and in these cases, the leading contributions arise instead from dimension-11 operators. For this reason, we consider the present analysis relevant and provide here a systematic study of such scenarios. The corresponding interactions can be traced back to the dimension-9 $|\Delta B| = 2$ LEFT operators introduced in Sec.~\ref{sec:eftframework}, which are not generated through tree-level matching from dimension-9 SMEFT operators, but rather by operators of dimension~11 or higher.

In general, all these operators arise from odd-dimensional operators with at least one $HH$ or $H^\dagger H$ insertion, and also possibly $n$ additional $H^\dagger H$ insertions. Consequently, the tree-level contribution to $n - \bar n$ coming from any of these operators will be proportional to $v \cdot (v/\Lambda)^{2n+1}$, where $\Lambda$ is another UV scale of the theory, and $v$ is the vacuum expectation value (vev) of the SM Higgs doublet, typically chosen to be $v \sim 176$ GeV. Among the three possible choices for $X_4$, we therefore exclude UV completions involving the scalar singlet $\mathcal{S}$, since its quartic coupling would, in the absence of an additional UV symmetry, generically permit the analogous trilinear interaction listed in Tab.~\ref{tab:DeltaB2models} without the $\mathcal{S}$ insertion. Such a trilinear term would dominate over the quartic one, making the phenomenology from the latter irrelevant.

For completeness, here we write the 22 UV models leading to dimension-11 $|\Delta B|=2$ SMEFT operators, distinguishing between those models involving two, three, or four exotic scalar multiplets. Interactions marked with \xmark\ correspond to models that, in the absence of an additional UV symmetry, would also permit one of the trilinear interactions listed in Tab.~\ref{tab:DeltaB2models}, , thereby inducing dimension-9 $|\Delta B|=2$ SMEFT operators. Each quartic interaction is implicitly accompanied by a dimensionless coupling $\lambda$.

\begin{enumerate}
    \item \textbf{Two exotic scalars:}
    This relatively simple model was noted earlier in Refs.~\cite{Helset:2021plg,Arnold:2012sd,Gardner:2018azu}, whose quartic interaction is given by
    \begin{equation} \label{eq:app1}
        \Upsilon \Upsilon \Upsilon \Xi_1 \, .
    \end{equation}
    
    \item \textbf{Three exotic scalars:} In this scenario we have two exotic scalars with diquark couplings in addition to $X_4 = \lbrace\Xi_0,\,\Xi_1\rbrace$, and we have listed seven models given by~\footnote{In Ref.~\cite{Gardner:2018azu} the authors also list in M14 the quartic coupling $\Upsilon\Upsilon\Omega_1 \Xi_1$, but we have checked that this coupling vanishes identically if there is only one copy of $\Upsilon$.}
    \begin{gather}
    \Xi_1\bar S_1 \bar S_1 \Upsilon^*, \quad \Omega_2 \Omega_2\Upsilon \Xi_1^* \text{\ \xmark}, \quad S_1 S_1\Upsilon^* \Xi_1^*, \quad S_1 S_3 S_3 \Xi_1^*, \\[4pt] S_3 S_3 \Upsilon^* \Xi_1^*, \quad
    \Omega_1 \Omega_1 \Upsilon \Xi_1, \quad S_3 S_3 \bar S_1 \Xi_0^* \, .
\label{eq:app2}
\end{gather}

    \item \textbf{Four exotic scalars:} This is the least minimal type of models, as in addition to $X_4 = \lbrace\Xi_0,\,\Xi_1\rbrace$ we have three distinct exotic scalars. We have listed 14 possibilities given by
    \begin{gather}
    \bar S_1 S_3 \Omega_2^* \Xi_1, \quad S_1 S_3 \Omega_1^* \Xi_1^*, \quad S_1 S_3 \Upsilon^* \Xi_1^*, \quad \bar S_1 S_3 \tilde S_1 \Xi_1^*, \quad S_3 \bar S_1 \Omega_4^* \Xi_1^* \text{\ \xmark}, \\[4pt] S_3 \tilde S_1 \Omega_2^* \Xi_1^* \text{\ \xmark}, \quad \bar S_1 \tilde S_1 \Upsilon^* \Xi_1^*, \quad \Omega_2 \Omega_4 \Upsilon \Xi_1 \text{\ \xmark}, \quad S_1 S_3 \bar S_1 \Xi_0^*, \quad S_1 S_3 \Omega_2^* \Xi_0^* \text{\ \xmark}, \\[4pt] S_1 \bar S_1 \Upsilon^* \Xi_0^*, \quad S_3 \bar S_1 \Omega_1^* \Xi_0^*, \quad S_3 \bar S_1 \Upsilon^* \Xi_0^*, \quad \Omega_1 \Omega_2 \Upsilon \Xi_0 \text{\ \xmark}\, .
\label{eq:app3}
\end{gather}

\end{enumerate}
To conclude, although one generally expects dimension-9 $|\Delta B|=2$ effective operators to dominate over higher-dimensional contributions, there exist models in which $|\Delta B|=2$ processes are generated via the quartic scalar topology rather than the cubic one, such as the SM extended by $\Upsilon$ and $\Xi_1$. Moreover, such scenarios provide a natural framework to accommodate Majorana neutrino masses through $\Xi_1$, while simultaneously forbidding $|\Delta B|=1$ processes, such as proton decay. In this topology, the contribution to $n-\bar n$ oscillations can be naively estimated as
\begin{equation}
    \delta m_n \sim \frac{\mu \, v^2}{M_{X_1}^2 M_{X_2}^2 M_{X_3}^2 M_{X_4}^2} \braket{\bar n |\mathcal{O}_i| n} \, .
\end{equation}
Defining a common BNV scale $\Lambda_{\rm BNV}$ as
\begin{equation}
    \Lambda_{\rm BNV}^7 \equiv \frac{M_{X_1}^2 M_{X_2}^2 M_{X_3}^2 M_{X_4}^2}{\mu} \, ,
\end{equation}
and assuming $\mathcal{O}(1)$ couplings in the UV, one obtains an approximate lower bound
\begin{equation}
    \Lambda_{\rm BNV} \gtrsim \mathcal{O}(10)\;\mathrm{TeV} \, ,
\end{equation}
which could potentially be probed at current and future colliders.

\bibliographystyle{JHEP}
\bibliography{main}

\end{document}